\documentclass[prb,twocolumn,superscriptaddress,floatfix,letterpaper]{revtex4}

\usepackage{graphicx,amssymb,bm,amsfonts,natbib,amsmath,graphics,color}

\usepackage[bookmarks=false,pdfstartview=FitH,colorlinks=true,citecolor=blue,linkcolor=blue]{hyperref}

\begin{document}

\title {A high-efficiency spin-resolved photoemission spectrometer combining time-of-flight spectroscopy with exchange-scattering polarimetry}

\author{C. Jozwiak} \email{cmjozwiak@lbl.gov} \affiliation{Advanced Light Source, Lawrence Berkeley National Laboratory, Berkeley, CA 94720, USA}\affiliation{Materials Sciences Division, Lawrence Berkeley National Laboratory, Berkeley, CA 94720, USA}\affiliation{Department of Physics, University of California Berkeley, CA 94720, USA}
\author{J. Graf} \affiliation{Materials Sciences Division, Lawrence Berkeley National Laboratory, Berkeley, CA 94720, USA}
\author{G. Lebedev} \affiliation{Advanced Light Source, Lawrence Berkeley National Laboratory, Berkeley, CA 94720, USA}
\author{N. Andresen} \affiliation{Advanced Light Source, Lawrence Berkeley National Laboratory, Berkeley, CA 94720, USA}
\author{A. K. Schmid} \affiliation{Materials Sciences Division, Lawrence Berkeley National Laboratory, Berkeley, CA 94720, USA}
\author{A. V. Fedorov} \affiliation{Advanced Light Source, Lawrence Berkeley National Laboratory, Berkeley, CA 94720, USA}
\author{F. El Gabaly} \affiliation{Materials Science Division, Lawrence Berkeley National Laboratory, Berkeley, CA 94720, USA}\affiliation{Sandia National Laboratories, Livermore, CA 94550, USA}
\author{W. Wan} \affiliation{Accelerator and Fusion Research Division, Lawrence Berkeley National Laboratory, Berkeley, CA 94720, USA}
\author{A. Lanzara} \email{alanzara@lbl.gov} \affiliation{Department of Physics, University of California Berkeley, CA 94720, USA} \affiliation{Materials Sciences Division, Lawrence Berkeley National Laboratory, Berkeley, CA 94720, USA} 
\author{Z. Hussain} \email{zhussain@lbl.gov} \affiliation{Advanced Light Source, Lawrence Berkeley National Laboratory, Berkeley, CA 94720, USA}

\date {\today}

\begin{abstract}
We describe a spin-resolved electron spectrometer capable of uniquely efficient and high energy resolution measurements.
Spin analysis is obtained through polarimetry based on low-energy exchange scattering from a ferromagnetic thin-film target.
This approach can achieve a similar analyzing power (Sherman function) as state-of-the-art Mott scattering polarimeters, but with as much as 100 times improved efficiency due to increased reflectivity.
Performance is further enhanced by integrating the polarimeter into a time-of-flight (TOF) based energy analysis scheme with a precise and flexible electrostatic lens system.
The parallel acquisition of a range of electron kinetic energies afforded by the TOF approach results in an order of magnitude (or more) increase in efficiency compared to hemispherical analyzers.
The lens system additionally features a 90$^{\circ}$ bandpass filter, which by removing unwanted parts of the photoelectron distribution allows the TOF technique to be performed at low electron drift energy and high energy resolution within a wide range of experimental parameters.
The spectrometer is ideally suited for high-resolution spin- and angle-resolved photoemission spectroscopy (spin-ARPES), and initial results are shown.
The TOF approach makes the spectrometer especially ideal for time-resolved spin-ARPES experiments.
\end{abstract}

\maketitle

\section{\label{sec:Intro}Introduction}

Understanding the complex roles and behaviors of electronic spin in matter has grown increasingly desirable as technological trends move towards materials and devices based on ever stricter control and utilization of the spin degree of freedom.
Combined spin- and angle-resolved photoemission spectroscopy (spin-ARPES) represents a powerful method of directly probing the spin physics of the energy and momentum dependent electronic systems of current research interests.\cite{Johnson1997}
While electron spectroscopies such as angle-resolved photoemission spectroscopy (ARPES) have been developed into precise, successful, and widespread techniques,\cite{hufner2003,Damascelli2003} spin-resolved versions comparatively suffer due to the large inefficiencies inherent to spin polarization measurements.\cite{Kesslerbook,Kirschnerbook,Federbook}
Despite the long-standing instrumental difficulties, numerous groups very recently have been re-illustrating the exciting potential of spin-ARPES for high impact research in the current science climate (e.g. Refs.~\onlinecite{Cinchetti2006,Varykhalov2008,Varykhalov2008a,Wells2009,Hsieh2009,Hsieh2009a,Dil2009,Qi2010}).
Works such as these add further urgency for developing improved spin-ARPES instrumentation.

The primary difficulty in such efforts is the lack of a direct way to measure the spin of free electrons.
Current polarimeters must function through various spin dependent effects in scattering electrons from a variety of targets and extracting their spin polarization as a differential intensity measurement.\cite{Kesslerbook,Kirschnerbook}
The most widely used polarimeters are variants of the Mott-polarimeter,\cite{Gay1992} in which the spin-orbit interaction introduces an asymmetry in the rate of backscattering into opposite directions dependent on the spin of the incident beam.
Unfortunately, the cross sections for these scattering events at the incident kinetic energies necessary for useful spin asymmetries (20-100 keV) are extremely low, resulting in several orders of magnitude decreased efficiency compared with straightforward spin-integrated intensity measurements.\cite{Gay1992}
Other types of polarimeters based on the spin-orbit interaction obtain similar performances\cite{Oepen1985,Unguris1986} (for a review of spin-orbit polarimeters, see Ref.~\onlinecite{Pierce1988}).

In general, spin-ARPES requires a spectrometer combining both a spin polarimeter and an electron energy (and angle) analyzer (EEA).
As much of the vast advancements in present high-resolution ARPES have been based on the popular hemispherical energy analyzers (HEA),\cite{Kevan1983,Maartensson1994} the most common spin-ARPES setup combines a Mott-polarimeter with an HEA.\cite{Huang1993,Fedorov1998,Ghiringhelli1999,Hoesch2002}
Indeed this is the type of spectrometer used for most of the recent high profile works.\cite{Varykhalov2008,Varykhalov2008a,Wells2009,Hsieh2009,Hsieh2009a}
The working principle of an HEA is the chromatic dispersion of an electron beam along a spatial dimension.
A significant part of the HEA's effectiveness is its compatibility with parallel 2-dimensional (2D) multi-channel detection along two spatial directions, corresponding to energy and one angular photoemission direction.\cite{Kevan1983,Maartensson1994}
However, as available spin polarimeters operate as effective single-spatial-channel detectors, the combined instrument loses this key characteristic and operates as a serial single-channel analyzer.
Compared to current spin-integrated HEA spectrometers, the Mott-HEA spectrometer is hindered by the single-channel acquisition of energy (and angle) itself, as well as the inherent inefficiency of the single energy (and angle) channel provided by the Mott-polarimeter.
Experiments are then forced to operate with significantly compromised energy and angle resolutions for adequate count rates.

We have developed a spectrometer at the Advanced Light Source (ALS) in Berkeley aimed at achieving a much improved ($>$100x) total efficiency compared to the Mott-HEA spectrometer, thus allowing spin-ARPES to be performed with the improved resolutions desired for current spin-dependent research.
This is achieved through two significant differences with respect to the Mott-HEA: (1) spin-resolution is provided by a higher efficiency low-energy exchange-scattering (LEX) polarimeter, and (2) energy analysis is performed with the time-of-flight (TOF) technique to achieve parallel multi-channel energy detection.
Two distinct modes of operation, a straight electron flight path and one which includes a 90$^{\circ}$ ``quarter-sphere'' bandpass filter (BPF), uniquely enhances the spectrometer's general utility as well as its ability to achieve high energy resolution with a wide range of light sources.

This paper is organized as follows.
Section II outlines the design concepts and operating principals of the new spin-ARPES spectrometer, including the TOF-EEA scheme, its integrated 90$^{\circ}$ BPF, the LEX polarimeter and its design optimizations, and the detection and data acquisition electronics.
Section III provides a detailed report of the spin-integrated performance of the TOF-EEA as well as initial spin-resolved measurements with the full spectrometer carried out at beamline 12.0.1 of the ALS operating in 2-bunch mode.
Section IV gives a concluding discussion.

\section{\label{sec:Design}Spin-resolved Time-of-Flight electron spectrometer design}

\begin{figure} \includegraphics[width=7cm]{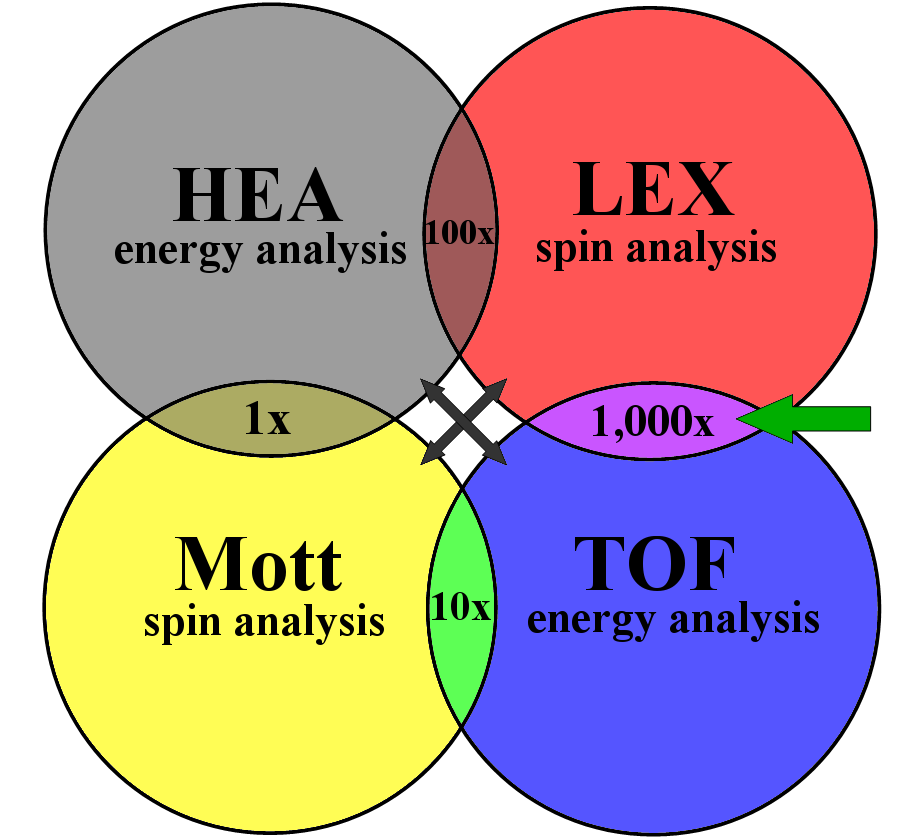}
\caption{\label{fig:ven}(color online) Schematic diagram depicting several designs of spin-ARPES spectrometers based on possible combinations of energy analyzers (HEA or TOF) and spin polarimeters (Mott or LEX).  The combinations are represented by the overlapping circles, and are labeled by the approximate efficiencies relative to the Mott-HEA combination (explained in text).  The combinations on the left, top, and bottom have been previously developed.  A spectrometer based on the combination on the right, marked by the green arrow, is reported here.}
\end{figure}

Although the Mott-HEA spectrometer combination seems to be the most popular, the strong desire for improved performance for spin-ARPES experiments has driven numerous and significant alternate approaches.
Most of these have involved the development of different combinations of EEAs and polarimeters, some of which are illustrated by the Ven diagram in Fig.~\ref{fig:ven}.
Each combination is represented by the overlap of the circles denoting the energy and spin alnalysis technique, with each overlap being labeled by approximate efficiency gains with respect to the Mott-HEA combination (which is therefore labeled \textit{1x}).

One general approach has been to replace the Mott-polarimeter with a LEX-type polarimeter.
Here we use the term to collectively describe polarimeter developments (e.g.~\onlinecite{Bertacco1999,Hillebrecht2002,Winkelmann2008,Okuda2008}) which utilize the spin-dependent specular reflectivity of ferromagnetic surfaces\cite{Waller1982,Alvarado1982,Kirschnerbook,Federbook} at energies of $\sim$10 eV and below.
The reflectivity's dependence on the relative alignment of the incoming electron's spin polarization and surface magnetization directions can be applied to spin polarimetry for nearly 100 times improved efficiency with respect to Mott-polarimeters.\cite{Fahsold1992,Hillebrecht2002,Okuda2008,Graf2005}
These improvements are due mainly to the higher average reflectivity of surfaces at such low energy compared to the total cross sections used at the higher energies of Mott-polarimeters.
In several of these cases, the given LEX-type polarimeter has been combined with an HEA\cite{Hillebrecht2002,Bertacco2002,Okuda2008} (or similarly with a cylindrical sector analyzer\cite{Winkelmann2008}) for performing spin-resolved experiments.
If electrons are efficiently transported from the HEA to the polarimeter in these cases, one expects a significant improvement ($\sim$100x) in total spectrometer efficiency.
Although the scattering target preparation and characterization is more involved than with Mott-polarimeters,\cite{Bertacco1999,Hillebrecht2002,Graf2005,Winkelmann2008,Okuda2008} the large improvement in effective count rate is well worth it for experiments requiring high resolution.
As discussed in Section~\ref{sec:SpinPol}, advances in passivation of the reactive target surfaces may aid the long term stability of a LEX polarimeter, improving its convenience.

Another alternative approach has been to keep the Mott-polarimeter, but replace the HEA with a TOF-EEA.
The TOF technique resolves photoelectron kinetic energy by measuring the time it takes to travel from its point of origin to final detection.
Excellent descriptions of the TOF technique and various TOF-EEAs are given in Refs.~\onlinecite{Bachrach1975,White1979,Hemmers1998}.
By moving energy resolution from the chromatic dispersion of electrons along a \textit{spatial} dimension (as in the HEA) to the \textit{temporal} dimension, multi-channel energy detection can be obtained even with effective single-spatial-channel polarimeters.
Recognizing the significant advantage of the parallel energy detection, several groups have developed successful spin-ARPES spectrometers based on integrating Mott-polarimeters into TOF-EEAs for use at BESSY,\cite{Mueller1995,Snell1999} the ALS,\cite{Snell2000,Snell2002} and more recently at the ESRF\cite{Moreschini2008} and Elettra.\cite{Cacho2009}
Specifically, comparing the efficiency of a single channel HEA with a TOF-EEA in acquiring spin-resolved data with the same energy resolution ($\Delta E$) across the same width energy window ($W$), the TOF approach can be a factor $W/\Delta E$ times more efficient, assuming similar electron transmission in both systems.\cite{woverde}
The larger this ratio, the greater is the benefit of the TOF-EEA.
A direct experimental comparison of the efficiencies of a Mott-TOF and Mott-HEA spectrometer is well presented in Ref.~\onlinecite{Moreschini2008}. 
The above mentioned groups using this type of spectrometer were interested in spin-polarized core levels and Auger lines and required measuring wide energy ranges which made the TOF approach quite suitable.
The primary focus of the current instrument is high energy resolution within a narrower energy window containing the low-energy physics near E$_F$; $W/\Delta E$ is still quite large in this case as $\Delta E$ is smaller.
For example, for studying a 1 eV energy window with 10 meV energy resolution, a TOF-EEA represents a 100-fold increase in data acquisition speed.

A significant caveat of the TOF technique for such ARPES experiments is the additional requirement of a pulsed light source with appropriate timing structure.
Third generation synchrotron light sources, such as the ALS, are pulsed sources with adequate time resolution, however usually operate at too high a repetition rate (e.g. 500 MHz at the ALS in ``multi-bunch mode'') for high resolution TOF measurements (see below).
Therefore, TOF-based spectrometers typically operate during periods of reduced repetition rate ($\sim$3 MHz at the ALS during ``2-bunch mode'') which allow high resolution TOF measurements, but translates to lower average photon flux ($\sim$10x at the ALS).
This must be taken into account when considering overall experiment efficiency.
In the example above, a TOF spectrometer used during 2-bunch mode at the ALS is then a factor of  $\sim$10 times more efficient than a single channel HEA spectrometer used during multi-bunch mode.
This is the source of the \textit{10x} label in the TOF-Mott overlap region in Fig.~\ref{fig:ven}. 
When comparing these approaches for use with light sources such as UV laser systems,\cite{Koralek2006,Meng2009,Graf2010} high-harmonic generation (HHG) laser systems,\cite{Haight1994,Chang1997,Siffalovic2001,Mathias2007} or free-electron lasers (FELs)\cite{Pietzsch2008} which may already have the required time structure, the issue of reduced flux is no longer a consideration, and the TOF-based spectrometer becomes even more ideal.
As advanced time-resolved pump-probe style experiments\cite{Cinchetti2006,Perfetti2006,Perfetti2007,Schmitt2008,Pietzsch2008} require such light sources and time structures independent of the spectrometer, they benefit from the full TOF efficiency gain.

While these different combinations have already been realized, Fig.~\ref{fig:ven} suggests that the most efficient spectrometer would be one which integrates a LEX-polarimeter within a TOF-based analyzer.
To the best of our knowledge, this approach has not been previously developed.
To make use of the large combined efficiency gains of this arrangement, we have developed a spin-resolved, TOF-based spectrometer (spin-TOF) with this approach.\cite{Jozwiakthesis}

\subsection{\label{sec:SpinPol}Low energy exchange-scattering spin polarimeter}

Electron spin polarimeters function by taking advantage of spin-dependent scattering processes and precisely measuring the magnitude of an asymmetry in the intensity of scattered electrons with respect to a particular scattering parameter.\cite{Kesslerbook,Kirschnerbook}
The polarization measurement is reduced to a differential intensity measurement in which the unknown polarization of the incident electron beam along a given axis, $P$, is proportional to the measured normalized intensity asymmetry, $A_m$, between two distinct scattering channels, $I_1$ and $I_2$.
The proportionality is generally expressed as
\begin{equation}\label{eqn:mainpol}
P = \frac{1}{S_{\textrm{eff}}}A_m \qquad \textrm{with} \qquad A_m = \frac{I_1-I_2}{I_1+I_2}
\end{equation}
where $S_{\textrm{eff}}$, the effective Sherman function, quantifies the analyzing power of the given scattering process and must be known beforehand in order to be used for polarimetry.

The statistical error in the polarization measurement can be simplified as\cite{Kesslerbook}
\begin{equation}\label{eqn:deltap}
\Delta P_{stat} = \frac{1}{S_{\textrm{eff}}}\Delta A_m \approx \frac{1}{\sqrt{S_{\textrm{eff}}^2N}} = \frac{1}{\sqrt{tI_0}}\frac{1}{\sqrt{S_{\textrm{eff}}^2\frac{I}{I_0}}}
\end{equation}
where $I$ is the total intensity of the beam measured in the two scattering channels, $I_0$ is the intensity of the incident electron beam, and $t$ is the total acquisition time (i.e. the total number of incident electrons is given by $N_0 = tI_0$).
In discussing the efficiency of a given polarimeter, one often defines the ``Figure of Merit'' $F$ as
\begin{equation}\label{eqn:fom}
F = S_{\textrm{eff}}^2\frac{I}{I_0}
\end{equation}
such that equation~\ref{eqn:deltap} can be rewritten as
\begin{equation}
\Delta P_{stat} \approx \frac{1}{\sqrt{tI_0}}\frac{1}{\sqrt{F}}.
\end{equation}
Expressed in this way, one can see that achieving a given statistical certainty in a polarization measurement requires a factor $1/F$ more time than a simple integrated intensity measurement of the same incident beam ($\Delta I_0 = 1/\sqrt{tI_0}$).
Therefore, a primary goal of polarimeter design is the maximization of $F$.

The systematic error in the polarization measurement due to uncertainty in $S_{\textrm{eff}}$ can be expressed as\cite{Getzlaff1998}
\begin{equation}\label{eq:syst}
\Delta P_{syst} = \frac{A_m}{S_{\textrm{eff}}^2}\Delta S_{\textrm{eff}}.
\end{equation}
Error due to instrumental intensity asymmetry that is unrelated to spin polarization\cite{Gay1992} can be approximately expressed as\cite{Bertacco1999}
\begin{equation}\label{eq:instr}
\Delta P_{instr} \approx \frac{A_{instr}}{S_{\textrm{eff}}}
\end{equation}
where $A_{instr}$ is the measured intensity asymmetry with an unpolarized incident beam.
These additional sources of error further emphasize the importance of a large Sherman function for spin polarimetry.
While additional measurements made by reversing the polarization of the source can sometimes be performed to aid in removing instrumental asymmetry from a polarization measurement,\cite{Gay1992,Johnson1997,Getzlaff1998,Ghiringhelli1999,Okuda2008,Bertacco1999} minimal $A_{instr}$ is preferred and essential for experiments which do not allow for the reversal of source polarization.\cite{Gay1992,Bertacco1999,Bertacco2002}

The most widely utilized mechanism for spin-dependent scattering in spin polarimetry has been the spin-orbit interaction,\cite{Kesslerbook,Kirschnerbook,Federbook} the most prominent example of which is the Mott polarimeter.\cite{Gay1992}
In a Mott polarimeter, sizable values of $S_{\textrm{eff}}$ are found at scattering energies between 20-100 keV where cross sections in general are low.
Hence, Mott polarimeters currently in use typically have values of $S_{\textrm{eff}}<0.20$ with total efficiencies of $F<2.5\times10^{-4}$.\cite{Pierce1988,Getzlaff1998,Gay1992,Burnett1994,Ghiringhelli1999,Petrov1997,Hoesch2002}
These low values of $F$ are a primary source of difficulty in spin-resolved experiments.
Mott polarimeters do allow a convenient target (typically Au foil) for a stable $S_{\textrm{eff}}$ over long time scales.
However they can also be quite sensitive to beam alignment, often leading to significant values of $A_{instr}$ which can be difficult to deal with.\cite{Gay1992}

Recently there has been a number of groups developing LEX spin-polarimeters that instead of spin-orbit coupling utilize the exchange interaction in low-energy scattering from ferromagnetic surfaces.\cite{Bertacco1999,Bertacco2002,Winkelmann2008,Okuda2008}
The exchange interaction introduces a dependence in the reflectivity of a target surface on the relative orientation of the polarization of the incident beam and the target's magnetization. 
These polarimeters make use of the resulting asymmetry in the intensity of the specularly reflected beam upon reversing the target's magnetization direction.
Equation~\ref{eqn:mainpol} is then applied with $I_1$ and $I_2$ being successive intensity measurements with the target magnetization reversed, and $P$ being the incident beam's polarization component along the target magnetization axis.
Measurement of an unknown polarization through this technique again requires prior knowledge or calibration of $S_{\textrm{eff}}$.
One of the exciting reasons for developing this type of polarimeter is that various target systems provide similar (or higher) values of $S_{\textrm{eff}}$ as Mott polarimeters, but at scattering energies $<20$ eV where the average scattering cross section is significantly higher.
Thus values of $I/I_0$ are much improved resulting in a larger $F$ (equation~\ref{eqn:fom}).
Specifically, reported values of $S_{\textrm{eff}}$ have ranged from $\sim$0.20\cite{Bertacco2002,Winkelmann2008} to $\sim$0.40\cite{Hillebrecht2002,Graf2005,Okuda2008} while those of $F$ have ranged from $7\times10^{-4}$\cite{Bertacco2002} (this value includes efficiency of coupling to the energy analyzer), to $2.2\times10^{-3}$\cite{Winkelmann2008}, to as high as $2\times10^{-2}$.\cite{Hillebrecht2002,Graf2005,Okuda2008} 

The strong exchange induced spin-dependence in low energy scattering from ferromagnetic surfaces had been proposed\cite{Feder1975,Feder1979} and experimentally studied\cite{Waller1982,Alvarado1982,Gradmann1983,Kirschner1984} several decades ago, and further suggested to be taken advantage of in efficient polarimeter development\cite{Tillmann1989,Hammond1990,Hammond1992,Fahsold1992} not long after.
A LEX-polarimeter achieving an efficiency of $F=2\times10^{-3}$ was indeed developed and coupled to a HEA for spin-resolved ARPES experiments as early as 1990.\cite{Hillebrecht1990}
The more recent increase of interest in LEX-polarimeter development\cite{Hillebrecht2002,Bertacco2002,Winkelmann2008,Okuda2008} is likely reflective of the current surge in importance of spin-resolved experimental capabilities.

\begin{figure} \includegraphics[width=8.5cm]{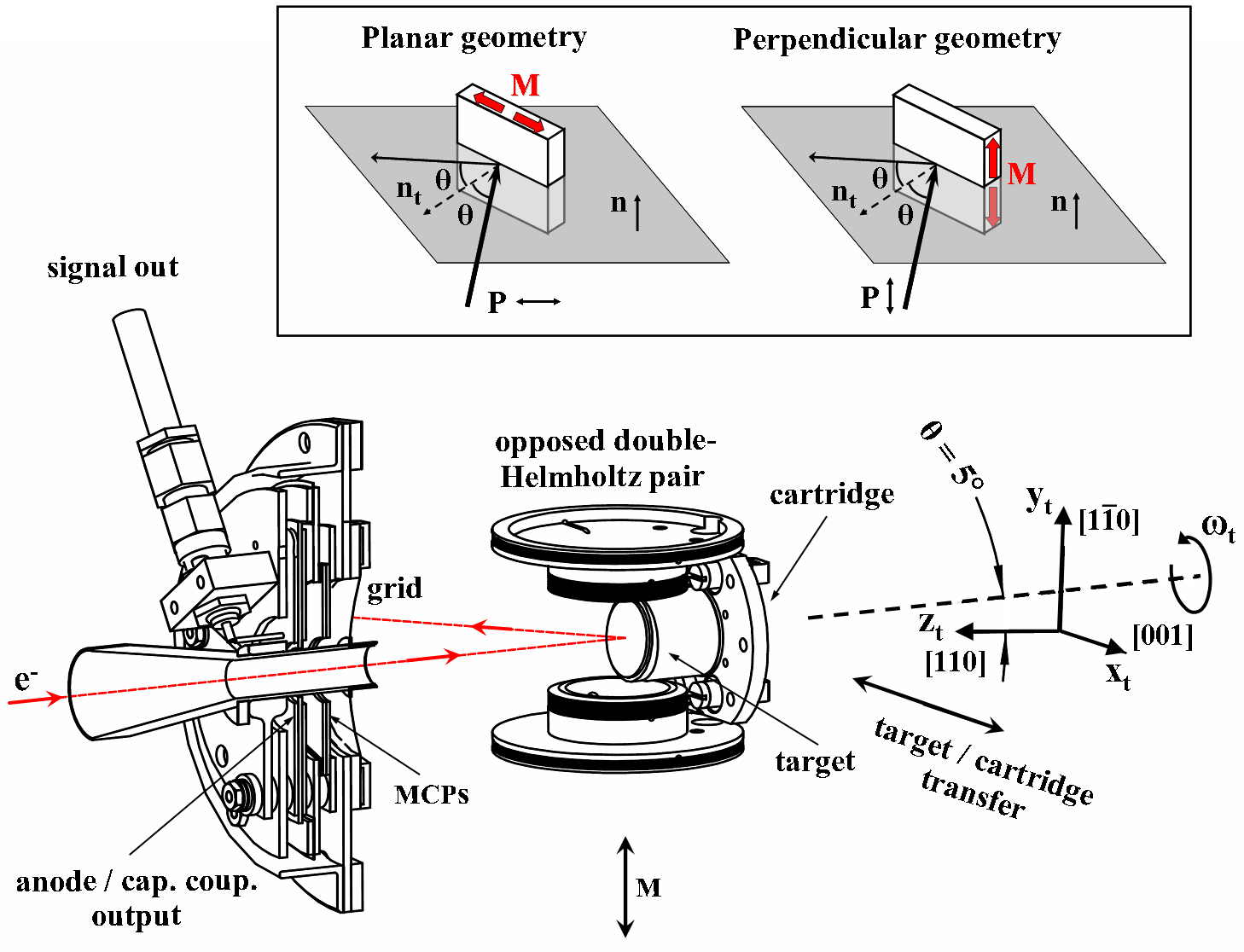}
\caption{\label{fig:SD}(color online) Drawing of the LEX polarimeter scattering and detection components.  MCP detector with annular tube is shown in sectioned view.  Electron path is shown by red dashed line.  The inset shows two useful scattering geometries; the main figure shows polarimeter configured in the Planar geometry with the scattering plane coplanar with the $\mathbf{y}_t\mathbf{z}_t$ plane.}
\end{figure}

In order to take advantage of the significantly higher values of $F$, we developed a unique version of the LEX-polarimeter capable of integration into a TOF-EEA.
In addition to fast time-resolution, the present design focuses on versatility to handle both a wide array of suitable scattering targets and experimental geometries.
The primary components are shown in Fig.~\ref{fig:SD}.
The electron beam to be spin analyzed enters a tube which passes through the center of a rear-facing multichannel plate (MCP) detector assembly and strikes the target surface, $\sim$50 mm downstream, at near normal incidence.
The specular (0,0) reflected beam is then collected by the front of the MCP detector, and fast output pulses marking the time of electron detection events are processed and recorded.
The scattering target's magnetization is switched in-situ by the coils along the target's ``y'' axis ($y_t$ in Fig.~\ref{fig:SD}); the polarimeter is then sensitive to the incident spin component along this axis.
A histogram of electron arrival times is built with the target magnetized in both directions.
The resulting asymmetry is then used to measure the spin polarization (via equation~\ref{eqn:mainpol}) of the incoming beam as a function of arrival time.
The time-resolution is required for the polarimeter's integration into a TOF-EEA (section~\ref{sec:TOFEEA}).

The electron detector utilizes a chevron pair of 27 mm effective OD multichannel plates (MCPs) with a 6 mm annular hole.\cite{Hamamatsu4294}
The MCP output is collected by a single anode disk, also with annular hole, coaxial with the MCPs, and the signal output is immediately capacitively decoupled by a second plate separated from the anode by a 0.005" Kapton film.
In contrast to channeltron detectors, the short charge transit distance through the 12 micron pores, $<$1.5 mm thick MCP stack, and signal output coupling scheme provide fast, $<$1 ns rise-time pulses capable of $<$ 200 ps time-resolution (discussed in section~\ref{sec:daq}).
The detector geometry is rotationally symmetric about the incident beam axis, allowing the target and scattering plane to be arbitrarily rotated by an angle $w_t$ (Fig.~\ref{fig:SD}), resulting in the reflected beam being recorded by a different but geometrically equivalent spot of the same detector.
The target assembly including the magnetization coils are mounted on a differentially pumped rotary platform to provide this rotation, thus enabling the polarimeter to be sensitive to spin along any axis within the plane normal to the incident beam, while keeping a fixed scattering geometry and therefore a constant $S_{\textrm{eff}}$.
This is distinct from the designs for multiple-axis sensitivity of other LEX polarimeters: the design of Okuda \textit{et al.}\cite{Okuda2008} provides two fixed axes accessed by separate detectors, while the design of Winkelmann \textit{et al.}\cite{Winkelmann2008} provides continuous rotation of the target and magnetization axis about the target surface normal, thus altering the scattering geometry and possibly $S_{\textrm{eff}}$. 

While LEX polarimeters take advantage of the exchange interaction, there are finite spin-orbit induced effects when scattering from heavy ferromagnets such as Gd,\cite{Weller1986} and even from light ferromagnets such as Fe.\cite{Waller1982,Kirschner1984,Bertacco1999}
Two scattering geometries, depicted in the inset to Fig.~\ref{fig:SD}, are often used for isolating and studying the exchange and spin-orbit interactions in scattering from crystals; the particulars of these geometries are well discussed in the literature.\cite{Ackermann1984,Kirschnerbook,Federbook}
It is not clear that there is a significant difference between the two in terms of polarimeter performance, however details of the two remain interesting.
Briefly, the `Planar' geometry, which is also shown utilized in the polarimeter schematic, has the target magnetization in the scattering plane, while the `Perpendicular' geometry has it perpendicular to the scattering plane.
Although other LEX polarimeters seem to use the latter, we have initially used the former as the exchange and spin-orbit interactions are more completely isolated.\cite{Kirschner1984,Kirschnerbook}
In this case the axis sensitive to exchange effects (the magnetization axis) is perpendicular to the axis sensitive to spin-orbit effects (the scattering plane normal, $\mathbf{n}$).
For example, in the Perpendicular geometry, interference effects between the spin-orbit and exchange interactions can lead to a measured asymmetry even with an unpolarized incident beam.
This can be pictured as the result of a double-scattering: in the first, there is a spin-orbit induced polarization along $\mathbf{n}$, followed by an exchange induced intensity asymmetry in the second.
This effect is removed in the Planar geometry.
One should note that spin-orbit effects are not completely removed in the Planar geometry, as interference can lead to the spin-orbit interaction first rotating the component of the incident polarization within the scattering plane, thus slightly altering the magnitude of the in-plane component's projection along the target magnetization.
As both effects are higher order, they should in general be small.
We also note that the present detector geometry allows a much smaller angle of incidence (only $5^{\circ}$ compared to other polarimeters at $15^{\circ}$\cite{Bertacco2002,Hillebrecht2002,Winkelmann2008,Okuda2008}) which may slightly improve $S_{\textrm{eff}}$ and $F$,\cite{Tamura1985,Bertacco1998} as well as reduce any spin-orbit effects even further (the spin-orbit interaction in the specular reflection must reduce to zero at normal incidence).
If necessary, it is straightforward to switch between the two geometries by simply changing the tilt direction of the target within the target and magnet coil assembly; again this flexibility is accommodated by the circular detector geometry.

The volume within the polarimeter, from the entrance tube to the scattering target, is maintained as a field free region.
As the LEX scattering is typically most efficient at kinetic energies below 10 eV, this region is very sensitive to magnetic fields.
Therefore, special attention was paid to ensure minimal magnetic fields through the use of double-walled magnetic shielding and the strict use of non-magnetic materials within the shields.
The magnetic target itself and its magnetizing coils present possible difficulties.
Using thin ferromagnetic films with in-plane magnetization direction grown on nonmagnetic substrates helps reduce stray fields from the target itself to a minimum.
The coils are a double set of Helmholtz coils (see Fig.~\ref{fig:SD}): the inner pair has half the radius and 4 times the number of turns as the outer pair, and are connected in series in opposing directions.
The field in the center at the target position due to the inner pair is 8 times stronger than the outer pair, resulting in a net field strong enough to reverse the magnetization of the target films ($\sim$100 G).
The approximate dipole fields far from the coil pairs, however, are the same magnitude and thus mostly cancel at the distance of the magnetic shielding.
These design aspects minimize changes in stray magnetic fields after magnetization switching that could lead to spurious $A_{instr}$ and error in polarization measurements (equation~\ref{eq:instr}).
Indeed the polarimeter measures an extremely small $A_{instr}$ (section~\ref{sec:spin}).

Of course the most important component of a LEX polarimeter is the target system itself.
As a number of systems have been used and suggested, and many more can be imagined, we have included a dedicated small and efficient multi-use preparation chamber capable of preparing a wide range of target surfaces and transferring them directly into the polarimeter in constant UHV.
The chamber includes a stage for target annealing and high temperature flashing, a quartz crystal monitor, an e-beam evaporator, and a precision gas leak valve.

For initial work, thin Co(0001) films grown on a W(110) substrate were used due to the potential of high values of $S_{\textrm{eff}}$ (0.40) and $F$ ($2\times10^{-2}$) of ultrathin films ($\sim$5 monolayers).\cite{Graf2005}
In general, W(110) is a straightforward substrate to clean for epitaxial film growth, and aged films can be quickly `flashed' off to start fresh.
It is also a good substrate for growing either Co(0001) or Fe(110) surfaces.\cite{Graf2005,Bansmann1997}
For the present Co films, substrate flash heating in O$_2$ and film growth proceeds as detailed in Ref.~\onlinecite{Bansmann1997,Bansmann1999} with a brief 30 s $\sim$500$^{\circ}$ C anneal.
The magnetic easy-axis is along the W $[1\overline{1}0]$ direction, thus the W crystal is oriented in the polarimeter as shown in Fig.~\ref{fig:SD}.
While the ultrathin films discussed in Ref.~\onlinecite{Graf2005} have much potential, they involve extra complications such as a rapidly varying $S_{\textrm{eff}}$ with energy and are quite sensitive to growth conditions.
It was found that thicker films of $\sim$50 monolayers proved to be more forgiving to growth procedures, gave more reliable results, and were sufficient for initial instrument commissioning.

\begin{figure} \includegraphics[width=8cm]{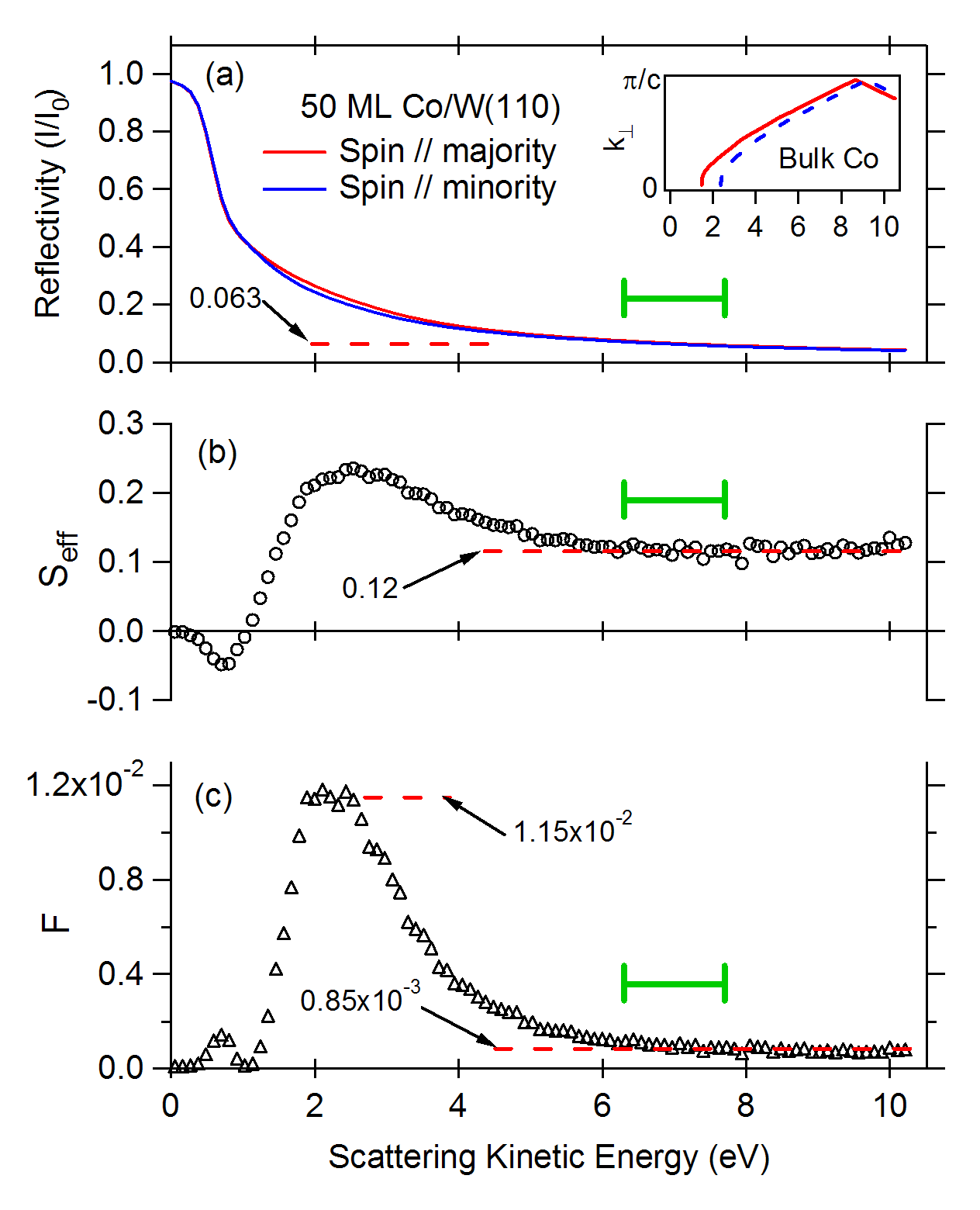}
\caption{\label{fig:spleem}(color online) Spin analyzing characteristics of a 50 monolayer Co/W(110) scattering target measured at normal incidence/reflection.  (a) Spin-dependent reflectivity with incident beam 20\% polarized along the target's magnetization direction.  Inset shows the spin-dependent band structure of bulk Co along the surface perpendicular direction (adapted from Ref.~\onlinecite{Scheunemann1997}).  (b) Corresponding $S_{\textrm{eff}}$.  (c) Corresponding $F$.  The green horizontal bars cover the approximate range of scattering energy used for the polarimeter data in this paper.}
\end{figure}

Similar to Ref.~\onlinecite{Graf2005}, 50 monlayer Co/W(110) targets were characterized with a Spin Polarized Low Energy Electron Microscope (SPLEEM),\cite{Bauer1994,Bauer2002} and the results are shown in Fig.~\ref{fig:spleem}.
Panel (a) shows the surface reflectivity of a single magnetic domain as a function of scattering energy with incident electrons 20\% polarized parallel and antiparallel to the Co majority spins.
The spin dependence is more clearly seen in the plot of $S_{\textrm{eff}}$ shown in panel (b), extracted as $S_{\textrm{eff}}=(1/P)(R_{\textrm{maj}}-R_{\textrm{min}})/(R_{\textrm{maj}}+R_{\textrm{min}})$, where $R_{\textrm{maj}}$ ($R_{\textrm{min}}$) is the reflectivity parallel to the Co majority (minority) spins, and $P$ is the polarization of the incident beam.
At $\sim$1 eV, there is a clear negative peak in $S_{\textrm{eff}}$ corresponding to the part of the Co bandstructure (Fig.~\ref{fig:spleem}) which has available majority states, but which is still in a minority bandgap which enhances the reflectivity of incident minority spins.
This is similar to the arguments for the observed behavior of Fe.\cite{Tamura1985,Hammond1990,Hammond1992,Fahsold1992,Bertacco1999,Hillebrecht2002,Winkelmann2008,Okuda2008}
At energies above the minority bandgap, the minority unoccupied density of states is then higher than the majority, which may be a reasonable explanation for the reversal of sign of $S_{\textrm{eff}}$.
There is a clear peak in $S_{\textrm{eff}}$ coincident with the bottom of the minority band, followed by a very flat region where there is minimal energy dependence of $S_{\textrm{eff}}$.
While the peak of 0.23 at 2.5 eV is appealing for maximum spin analyzing power, the initial work in the present paper is performed using scattering energies in the range highlighted by the horizontal green bars in Fig.~\ref{fig:spleem} due to the relative simplicity of an energy independent $S_{\textrm{eff}}$ (0.12) and higher scattering energies.
Fig.~\ref{fig:spleem} (c) displays the calculated $F$ from panels (a) and (b).
The peak value for this target system is quite high ($>1\times10^{-2}$), however the work in the present paper was performed at scattering energies with an $F$ just less than $1\times10^{-3}$, which is still significantly higher than the best Mott polarimeters.

It is possible that an Fe(001) surface will give increased performance in the current polarimeter, as previous LEX polarimeters have shown great performance with this surface.\cite{Bertacco1999,Hillebrecht2002,Winkelmann2008,Okuda2008}
As the exchange split bandgap of this surface also falls at a higher kinetic energy (6-10 eV) than the present Co(0001) surface, the related peaks in $S_{\textrm{eff}}$ and $F$ fall at these higher kinetic energies and are likely more convenient to use.
Additionally, the passivation of the Fe surface through oxidation offers enhanced stability and lifetimes.\cite{Bertacco1999,Winkelmann2008,Okuda2008}
The versatility of the dedicated target preparation system allows straightforward adaptability to this target surface and others, including those that have not been thoroughly explored yet.
For example, passivating ferromagnetic surfaces with a capping layer of graphene for substantially improved target lifetimes\cite{Dedkov2008,Dedkov2008a} may be an attractive option, and could be performed in the present preparation system.

The described LEX polarimeter could provide a variety of experimental systems with enhanced performance and efficiency with respect to more traditional Mott polarimeters.
In the particular application of spin-ARPES where energy and angular resolution are also required, we have fully integrated the polarimeter into the multifunction TOF-EEA described below.

\subsection{\label{sec:TOFEEA}TOF electron energy analyzer with 90$^{\circ}$ bandpass filter}

\begin{figure*} \includegraphics[width=18cm]{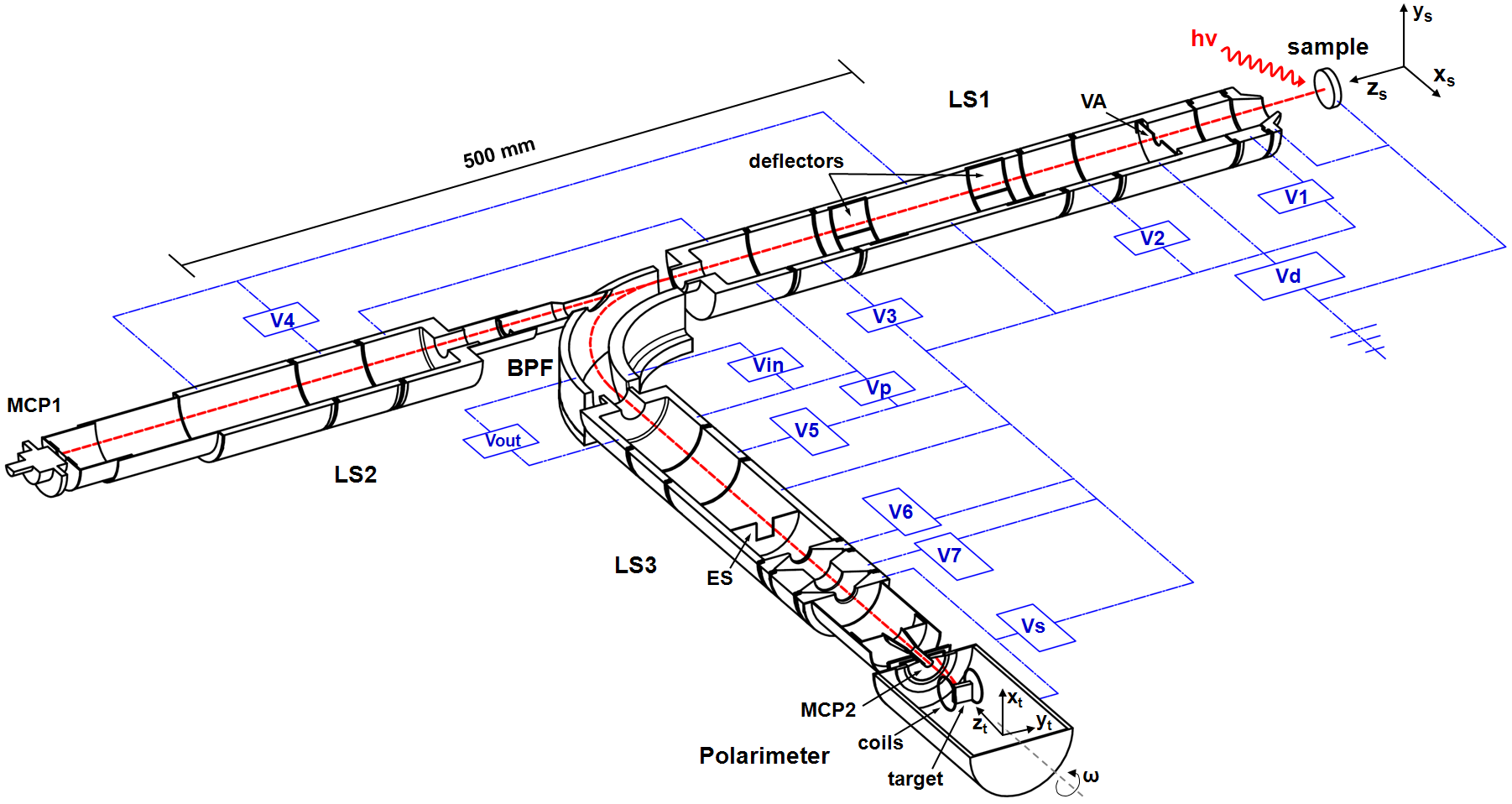}
\caption{\label{fig:main}(color online). Diagram of the spin-TOF spectrometer.  LS1, LS2, LS3: Lens System 1, 2, and 3.  BPF: Bandpass filter.  VA: variable aperture.  ES: exit slit.  Each electrode is wired as shown by 12 independent voltages.  Deflectors are powered by 4 additional independent voltages (not shown).  Electron paths for mode 1 (straight) and mode 2 (90$^{\circ}$ bend) are shown by the dashed red line.}
\end{figure*}

Commercially available HEAs are often the starting point in developing Mott-HEA spectrometers.
Indeed, full Mott-HEA spectrometers are now commercially available.
TOF-EEAs are less common, however, and one with the characteristics required for the present project and compatible for use with the ALS 2-bunch mode light specifications ($\sim$3 MHz repetition rate) required a fully custom design.
Some of the electron optical considerations of the resulting design which are not covered here are discussed in Ref.~\onlinecite{Lebedev2008}.

A cutaway diagram of the full spin-TOF spectrometer is shown in Fig.~\ref{fig:main}, illustrating the basic layout of the resulting TOF-EEA.
The design includes two modes of operation with distinct electron flight paths and detectors.
In the first mode (M1), electrons photoemitted from a sample by pulsed photons (far right) enter the first electrostatic lens system (LS1) and are transported along a nominally straight path through to a second lens system (LS2) where they are finally detected by a time-resolved detector (far left).
In the second mode (M2), photoelectrons exiting LS1 are electrostatically deflected 90$^{\circ}$ by a nested pair of spherical electrodes and transported through a third lens system (LS3) and focused onto another time-resolved detector (bottom).
As only photoelectrons of energies within a certain range around a nominal energy (E$_{pass}$) are successfully passed through the 90$^{\circ}$ bend, it acts as a bandpass filter (BPF), allowing only the desired section of a photoelectron spectrum to be detected in this mode.
A downstream variable exit slit (ES) aids in the removal of scattered and secondary electrons generated in the BPF.
A variable entrance aperture (VA) at the front of LS1 (with a selectable diameter of 2, 4, 6, or 8 mm) defines the angular acceptance, and therefore the angular resolution, of the spectrometer which ranges from $\sim \pm 0.5^{\circ}$ to $\sim\pm 2^{\circ}$.
Two sets of quadrupole deflectors within LS1 allow for corrections of the photoelectron beam path due to misalignment of the photon beam and sample with respect to the spectrometer or small stray magnetic fields.

The two distinct modes provide convenient flexibility in measurements.
Although detectors can be swapped, the primary configuration has a simple, single-anode high-speed MCP assembly for spin-integrated measurements at the end of LS2 for M1 operation, and the LEX-polarimeter (section~\ref{sec:SpinPol}) for spin-resolved measurements at the end of LS3 for M2 operation.
As we will see below, M2 operation with the 90$^{\circ}$ BPF (along with the polarimeter) provides for efficient, high energy resolution measurements over a certain range of the photoemission spectrum and with a wide range of photon energies.
Even with the high efficiency LEX-polarimeter, such spin-resolved measurements are more time consuming than straightforward spin-integrating detection.
Thus the ability to operate in M1 for rapid, lower resolution (higher flight energy), full spectrum acquisition of spin-integrated data is very convenient for system alignment, initial characterization and crystalographic alignment of the sample under investigation, and identifying sections of the spectrum where high energy- and spin-resolution is desired.
Switching between the two modes of operation takes a matter of seconds.

As the goal of the project is to provide high energy resolution for efficient and practical spin-resolved measurements, a primary design requirement of the TOF system is the capability of $\sim$10 meV resolution.
Thorough analyses of energy resolution in a TOF experiment are found in Refs.~\onlinecite{Hemmers1998,Samarin2003}.
Briefly, the kinetic energy in a field-free TOF measurement is related to both the electron's time of flight from photoemission to detection ($t$) and the length of its flight path ($L$) through
\begin{equation}\label{eq:en}
E = \frac{1}{2}m \frac{L^2}{t^2}.
\end{equation}
Straightforward differentiation of eq.~\ref{eq:en} shows that the energy resolution ($\Delta E$) has a contribution due to the total timing resolution ($\Delta t$) of
\begin{equation}\label{eq:dt}
\Delta E_t = \frac{2}{L}\sqrt{\frac{2}{m}} E^{3/2} \Delta t.
\end{equation}
It is clear that $\Delta E_t$ is dependent on the flight energy.
Thus the electrostatic lens columns are utilized to retard the photoemission spectrum to lower flight energies for high energy resolution, largely independent of initial photoelectron kinetic energy (i.e. photon energy) while maintaining high transmission for the portion of the spectrum of interest (e.g. Refs.~\onlinecite{Hemmers1998,Moreschini2008}).
This is in general reasonably practical with retardation ratios (the ratio of initial photoelectron kinetic energy to retarded average flight energy) of on the order of 10 or less.
One should note that for use with low photon energy laser sources, the photoelectrons are emitted at such low kinetic energies that the lens columns can instead be used to slightly accelerate the spectrum to more manageable flight energies.

The total timing resolution can be written as a combination of the excitation source pulse width ($\Delta t_{source}$), the resolution of the final detector ($\Delta t_{det}$), the resolution of the assorted electronics and data acquisition components used ($\Delta t_{daq}$), and any timing aberrations along the accepted flight paths of the EEA ($\Delta t_{EEA}$) as $\Delta t = \sqrt{\Delta t_{source}^2 + \Delta t_{det}^2 + \Delta t_{daq}^2 + \Delta t_{EEA}^2}$.
In the case of current synchrotron light sources, such as the ALS, the dominant contribution comes from a $\Delta t_{source}$ of around 100 ps; it is relatively straightforward to ensure the detector and electronics provide similar performance such that the total $\Delta t<200$ ps.
$\Delta t_{EEA}$, in part due to differences in flight path lengths, can be kept to a minimum ($<$50 ps) in the present TOF-EEA design such that they do not significantly add to the overall effective $\Delta t$.\cite{Lebedev2008} 
While energy resolution is inversely proportional to flight length, space and other constraints limited the size of the instrument to a reasonable choice of $L\sim1$m.
From equation~\ref{eq:dt} we then have $\Delta E_t \sim 2.4\times 10^{-4} E^{3/2}$, with $E$ and $\Delta E$ in eV.
To achieve 10 meV total resolution, one is limited by $\Delta E_t$ to work with flight energies of less than 12 eV.

\begin{figure} \includegraphics[width=8cm]{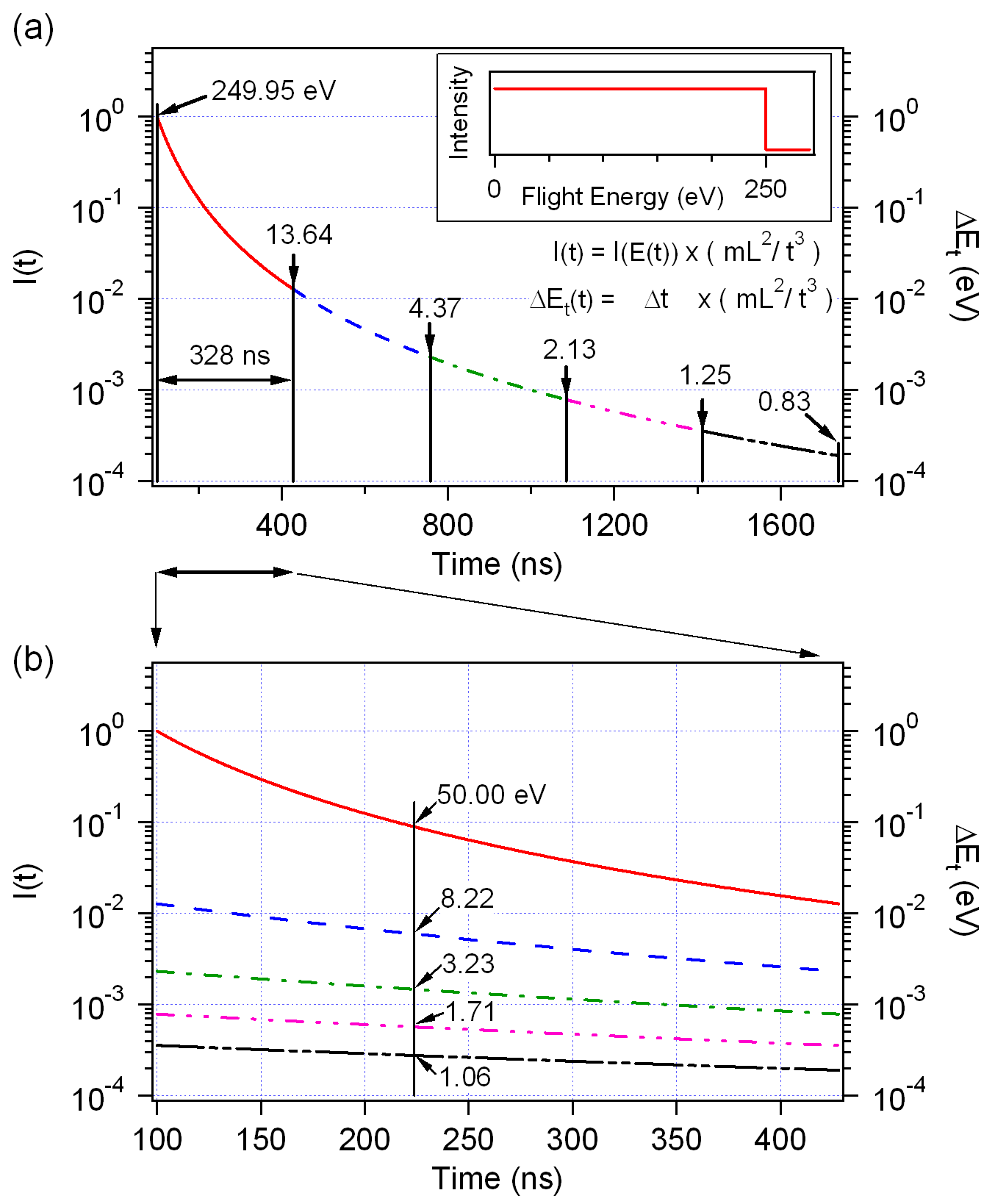}
\caption{\label{fig:overlap}(color online) (a) Schematic TOF spectrum, $I(t)$, which would result from the evenly distributed energy spectrum shown in the inset and with a flight length of 0.937 m.  The corresponding energy resolution, $\Delta E_t$, is shown by the right side axis assuming $\Delta t=200$ ps.  Vertical black lines separate the spectrum into 328 ns wide segments.  Flight energies corresponding to the flight times at these locations are marked.  (b) With a 328 ns period between light pulses, the effective TOF spectrum is the sum of the segments in panel (a), shifted into the same time window and overlapping as shown.  The vertical black line highlights relative intensities (resolutions) on the left (right) axis in the 220 ns time channel corresponding to the different flight energies labeled.}
\end{figure}

An important consideration is the possible overlap of lower (higher) energy electrons created by earlier (later) photon pulses reaching the detector at the same time.
This temporal overlap of a photoemission spectrum greatly complicates the extraction of real energy spectra, and needs to be avoided.
The problem is illustrated in Fig.~\ref{fig:overlap} with the example of the straight flight path of M1 (total flight length of 0.937 m) and the ALS 2-bunch mode (328 ns period).
The curve in panel (a), I(t), represents the relative number of electrons arriving in fixed width time channels as a function of time assuming a perfectly flat distribution of photoelectrons in energy.
The strong time dependence of this curve (note the log scale) is due to the non-linear conversion between time and energy scales (equation~\ref{eq:en}).
When converting from an energy to a time distribution, one must scale the intensities by the Jacobian, $dE/dt$; similarly for the real experiment, when converting a spectrum measured as a histogram in time channels to an energy histogram in energy channels, one scales the intensities by $dt/dE$.\cite{Trevor1989,Hemmers1998,Samarin2003}
Or, more precisely,
\begin{eqnarray}
I(t) & = & I(E(t)) \times \frac{dE}{dt}  =  I(E(t)) \times \frac{mL^2}{t^3} \quad \mathrm{and}\\
I(E) & = & I(t(E)) \times \frac{dt}{dE}  =  I(t(E)) \times L \sqrt{\frac{m}{(2E)^3}}.
\end{eqnarray} 
The curve in Fig.~\ref{fig:overlap} also represents the energy resolution, $\Delta E_t$ (right hand vertical axis), as it of course has the same functional time dependence through $\frac{dE}{dt}$.

One can set the available 328 ns time window to start at $t=100$ ns to coincide with the arrival of the fastest photoelectrons in this example (250 eV).
The vertical black bars in panel (a) mark the times of successive photon pulses with the corresponding flight energies as labeled.
As the detector cannot distinguish electrons which arrive at time $t$ from those at time $t+(n\times328)$ ns (where $n$ is an integer), the measured time spectrum will be the sum of each part of the curve in the upper panel separated by the vertical black lines.
This is shown in panel (b) where each section is translated back ($n\times328$) ns; the measured spectrum is the sum of each of these curves.
This shows the relative weight of counts at a given time channel originating from the different possible energies.
For example, the total intensity measured at $t=220$ ns is made up of 89.3\% 50 eV electrons, 6.0\% 8.22 eV, 1.5\% 3.23 eV electrons, etc.
Thus, if one is interested in studying part of a spectrum with a flight energy of 50 eV, the measured time spectrum will be noticeably overlapped by photoelectrons at flight energy $\sim$8 eV and 3 eV.
Further, if the section of interest of the photoemitted spectrum is transported at lower flight energies (possibly after retardation for improved resolution), the situation is worse.
If the electrons of interest are measured at 8.22 eV flight energy, for example, the ratio of desired intensity at 220 ns to that of lower energy electrons is further reduced.
More severe is the overlap with higher energy photoelectrons (if present in the originating spectrum) with a flight energy of 50 eV, as they will be proportionally much more intense in the time spectrum.
Even if one is interested in photoelectrons near the Fermi level, the much smaller number of photoelectrons at higher energies due to higher-order light at synchrotron sources can cause serious overlap problems (see Fig.~\ref{fig:enres} and discussion in section~\ref{sec:energy}).

It is interesting to note that overlap problems are less restrictive using a shorter flight path.
From equation~\ref{eq:en}, reducing $L$ decreases the lowest energy, $E$, which will arrive at the detector within the first 328 ns.
Electrostatic retardation will need to be increased accordingly to maintain the same energy resolution (equation~\ref{eq:dt}).
While this approach may reduce spectral overlap issues, the increased retardation required will result in decreased electron transmission and overall efficiency, which could be prohibitive for low count rate experiments such as spin-resolved measurements.

Spectral overlap can be completely removed with the incorporation of spatially dispersive elements, such as the 90$^{\circ}$ BPF (Fig.~\ref{fig:main}), to remove photoelectrons of unwanted flight energies.
In the example of Fig.~\ref{fig:overlap}, if one is interested in studying a portion of the photoemission spectrum with a flight energy of $\sim$ 8 eV for high energy resolution, one can utilize M2.
The BPF will pass only electrons within an energy range about the desired $E_{\textrm{pass}}$; any overlapping electrons at flight energies of 3 and 50 eV are not transmitted to the final detector.

While the spatial dispersion of electrons within the BPF is used to coarsely filter the spectrum, the temporal dispersion through the entire flight path is still used for parallel detection and resolution of electron energies within the passed energy window.
The present design provides a typical bandpass of $\sim$ 1 eV.
As discussed above, parallel detection with $\Delta E$=10 meV across an energy window of $W$=1 eV represents a $W/\Delta E$=100 times increase in efficiency with respect to serial detection, assuming equal electron transmission and photon flux.
In the context of high resolution spin-ARPES measurements, such as for the study of strongly correlated electron systems (e.g.~Refs.~\onlinecite{Damascelli2003,Qi2010,Hsieh2009,Hsieh2009a,Wells2009}), acquisition of a 1 eV window around E$_F$ or some other particular spectral feature is all that is typically required.
For taking spectra of wider energy ranges, the voltages of the TOF-EEA can be incrementally adjusted to `sweep' the window passed by the BPF across the energy range desired.
Thus, the TOF-EEA with the integrated dispersive BPF is a powerful combination for performing high resolution (low flight energy) TOF spectroscopy with light sources of higher energy and/or repetition rates than would otherwise be possible.
For example, this analyzer allows for high-resolution spectra to be taken with typical synchrotron light sources (and/or laser sources) with photon energies ranging from $\sim$6 eV to as high as several hundred eV and repetition rates as high as $\sim$10 MHz.

The 90$^{\circ}$ BPF offers additional benefits in the context of spin-resolved measurements.
As discussed in Section~\ref{sec:SpinPol}, the LEX-polarimeter is sensitive to spin direction along the scattering target's magnetization direction, or the $y_t$ direction as drawn in Fig.~\ref{fig:main}.
Photoelectron spin direction is not rotated through the 90$^{\circ}$ bend of the BPF, so the polarimeter is thus sensitive to the out-of-plane polarization component of the photoemitting sample, $z_s$.
By rotating the scattering target within the polarimeter by 90$^{\circ}$ about $\omega_t$, the polarimeter becomes sensitive to the polarization component along the sample's vertical in-plane axis, $y_s$.
Rotating the entire spectrometer 90$^{\circ}$ about the optical axis of LS1/LS2 aligns the polarimeter to access components along the sample's horizontal in-plane axis, $x_s$ (as well as still the $z_s$ axis).
Thus the 90$^{\circ}$ BPF allows measurements of spin polarization along all three sample axes if desired. 

The 90$^{\circ}$ BPF adds a number of complications, as well.
The dispersive and focal properties of the spherical electrodes result in finite timing aberration contributions ($\Delta t_{EEA}$) as electrons take flight paths of differing lengths through the orbit depending on energy, entrance angle, and entrance position.\cite{Imhof1976,Lower1989}
These timing aberrations decrease with increasing $E_{\textrm{pass}}$ (as $1/\sqrt{E_{\textrm{pass}}}$ for 180$^{\circ}$ sectors\cite{Imhof1976}).
As increasing $E_{\textrm{pass}}$ gives larger passed energy windows, it is in general advantageous to use the BPF with as high $E_{\textrm{pass}}$ as possible ($\sim$ 50 eV in practice).
For typical synchrotron photon energies, this means retarding photoelectrons through LS1, accelerating them for a high $E_{\textrm{pass}}$ through the BPF, and then retarding them again in LS3 for a low average flight energy.
By focusing the electron beam at the midpoint of the BPF (45$^{\circ}$), instead of at the entrance as is typical of 180$^{\circ}$ HEAs, the path length dependence on BPF entrance angle is removed, thus significantly reducing timing aberrations even for a large initial acceptance angle into the spectrometer.
This effect on timing aberrations can be seen by tuning the BPF focal point with careful adjustment of the voltage $V3$ (Fig.~\ref{fig:main}) to a minimum in observed energy/timing resolution during an experiment.
Remaining timing aberrations due to positional dependence, or focal spot size, in the BPF can be kept insignificantly low with a small source spot; the 100 micron photon spot sizes available at current synchrotron light sources is sufficiently small. 
Our calculations show that with such a spot size and proper tuning, BPF timing aberrations for a wide range of lens system parameters is kept below 15 ps (much less than $\Delta t_{source}$ and $\Delta t_{det}$), and is thus only a small contribution to total energy resolution.

\begin{figure} \includegraphics[width=7cm]{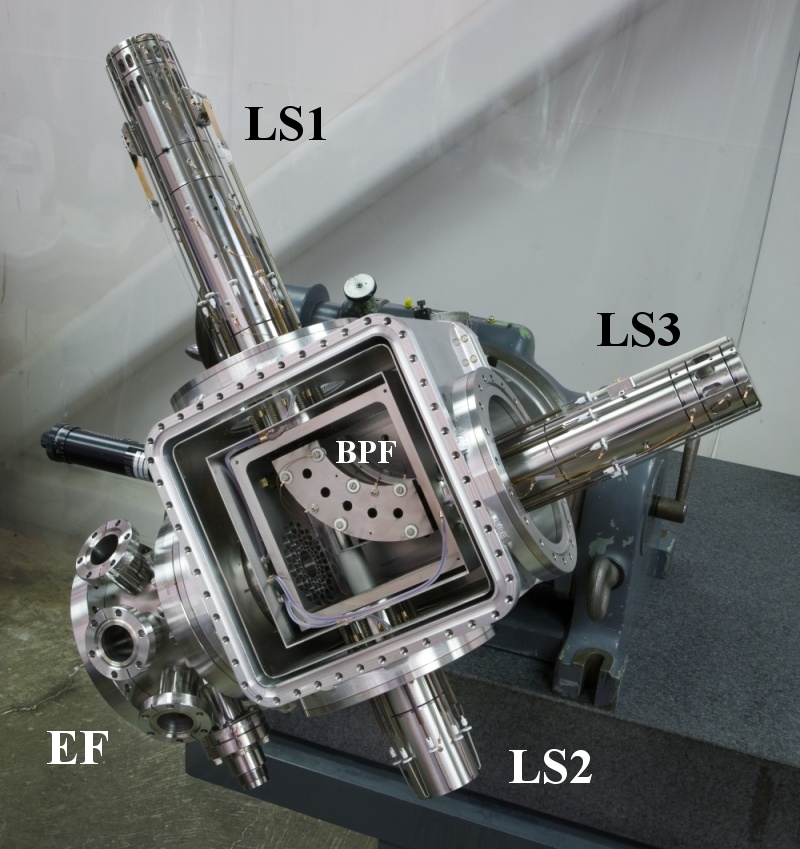}
\caption{\label{fig:pic}(color online) Photograph of the TOF lens system.  Magnetic shielding and vacuum enclosures removed for view.  EF: electrical feedthru ports.}
\end{figure}

Each electrode in the full lens system is titanium, coated with colloidal graphite.
Each lens column is a stack of cylindrical electrodes, electrically isolated by rings of 52 sapphire balls (1/8'' diameter), secured by 8 axial springs, and cantilever mounted from a central `box' electrode which houses the BPF.
The design and high machining tolerances limited radial displacement of electrodes in each assembled column to less than 32 micron from their central axes.
When the full assembly is rotated such that a lens column moves from a vertical to horizontal position, the ends of LS1, LS2, and LS3 sag a total of 34, 3, and 10 microns, respectively.
This small degree of electrode misalignment is calculated to contribute less than 5 ps of temporal aberrations to the overall $\Delta t_{EEA}$.
Full double-layered magnetic shielding surrounding the entire flight paths and all non-magnetic construction within keep stray magnetic fields along the flight paths to $<0.5$ mG.
A photograph of the assembled TOF lens system, without magnetic shielding, vacuum chambers, and detectors, is shown in Fig.~\ref{fig:pic}.

\subsection{\label{sec:daq}Data Acquisition}

\begin{figure} \includegraphics[width=8.5cm]{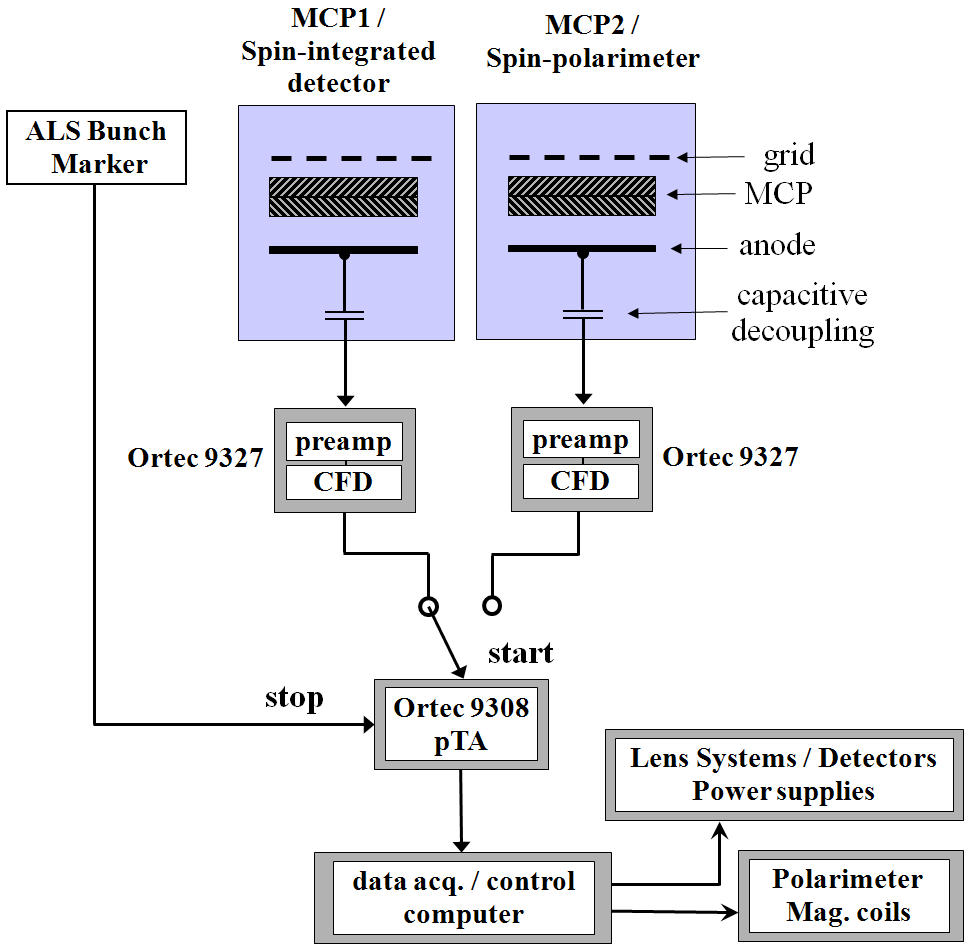}
\caption{\label{fig:elec}(color online) Schematic of data acquisition and instrument control electronics.}
\end{figure}

The simple signal processing and data acquisition scheme is shown in Fig.~\ref{fig:elec}.
In M1 use, an electron striking the MCP1 assembly\cite{Hamamatsu4655} (Fig.~\ref{fig:main}) generates a negative output pulse with a rise time of $\sim$415 ps through an in-vacuum 150 pF decoupling capacitor.
In M2 use with the LEX-polarimeter, the MCP2 assembly (Fig.~\ref{fig:main}) generates a negative output pulse with a rise time of $\sim$525 ps through a custom, on-board capacitive decoupling of 300 pF.
In either case, the signal is directly fed into dedicated electronics packages (Ortec, model 9327) that combine a 1-GHz preamplifier and constant fraction discriminator (CFD) that outputs a NIM-style digital pulse marking the time of signal input with a jitter and walk of $<20$ ps and $<40$ ps, respectively.
These signals are then fed into the Ortec 9308 Picosecond Time Analyzer (pTA), along with a timing signal marking the period of the pulsed light source, in this case the ALS beamline bunch marker.
The pTA builds the time-of-flight histogram in 65,536 time bins over a variable window from 80 ns to 163.84 $\mu$s; to match the period of the ALS 2-bunch mode, this is set to 320 ns providing a digital resolution of 4.88 ps.
The pTA is controlled and accessed through a direct connection to a desktop computer through a supplied PCI-card.
The same computer controls the power supply bank (from VG Scienta) which supplies the desired voltages to the spectrometer electrodes and detectors, as well as controls the polarimeter target magnetization coils.
Software written in LabVIEW (National Instruments) integrates control over spectrometer operation, data acquisition, and conversion of the time-of-flight histogram to an energy spectrum.\cite{Hemmers1998}
Data acquisition is easily synchronized with magnetization pulses such that spectra can be built up over frequent ($\sim$each minute) magnetization reversals to remove any effects of changing sample, target, or photo flux characteristics over time from the polarization measurement.

Switching between detectors involves only connecting the BNC cable from the desired preamp/CFD to the pTA input.
Note that the detector signals are used as the `Start' and the photon signal as the `Stop' such that the recorded time spectrum is reversed, as is often done in TOF spectroscopy.\cite{Moreschini2008,Vredenborg2008,Hemmers1998,White1979}
This reduces possible signal loss due to unnecessary dead time in the pTA as the expected count rate is far less than 1 count for every light pulse.\cite{OrtecAppNote52}

During an experiment, it is possible to record the signal from specularly reflected photons from the photoemitting sample.
This `photon peak' not only serves as a prompt for $t=0$ in the TOF spectra,\cite{Bachrach1975,Hemmers1998} but its width also represents the total timing resolution, $\Delta t$, of the entire setup.
We have directly measured with the ALS 2-bunch a $\Delta t$ of 137 ps and 195 ps for the spin-integrating MCP1 and the LEX-polarimeter MCP2, respectively.
Although more sophisticated TOF setups have recently achieved timing resolutions of 18 ps,\cite{Vredenborg2008} the present system is fundamentally limited by the $\sim$100 ps bunch width.\cite{timingcomment}
Thus the simple scheme used here (Fig.~\ref{fig:elec}), which involves only 3 commercial electronics packages and 6 total cables from the detectors to computer, is more than sufficient for high energy resolution in the present spectrometer, and is surprisingly staightforward and convenient.

\section{\label{sec:perf}Performance}

Initial experiments have been completed for demonstration and characterization of the spectrometer and its performance.
Data was taken at beamline 12.0.1.1 of the ALS during 2-bunch mode.
Photoemission was performed on a W(110) crystal as it provides a well known spectrum including sharp and easily accessible core levels for convenient energy references, and provides strong signal at a sharp Fermi edge at low temperature for energy resolution measurements.
It also serves as a substrate for evaporating high quality Au(111) films\cite{Bauer1977,Shikin2002} used for demonstrating the polarimeter performance.
The crystal was mounted with the [1$\overline{1}$0] direction oriented along the vertical $y_s$ axis (Fig.~\ref{fig:main}); the crystal is rotated about the vertical polar axis to change the emission angle accepted by the spectrometer along the [00$\overline{1}$] direction.

\subsection{\label{sec:energy}Energy calibration and resolution}

\begin{figure} \includegraphics[width=8.5cm]{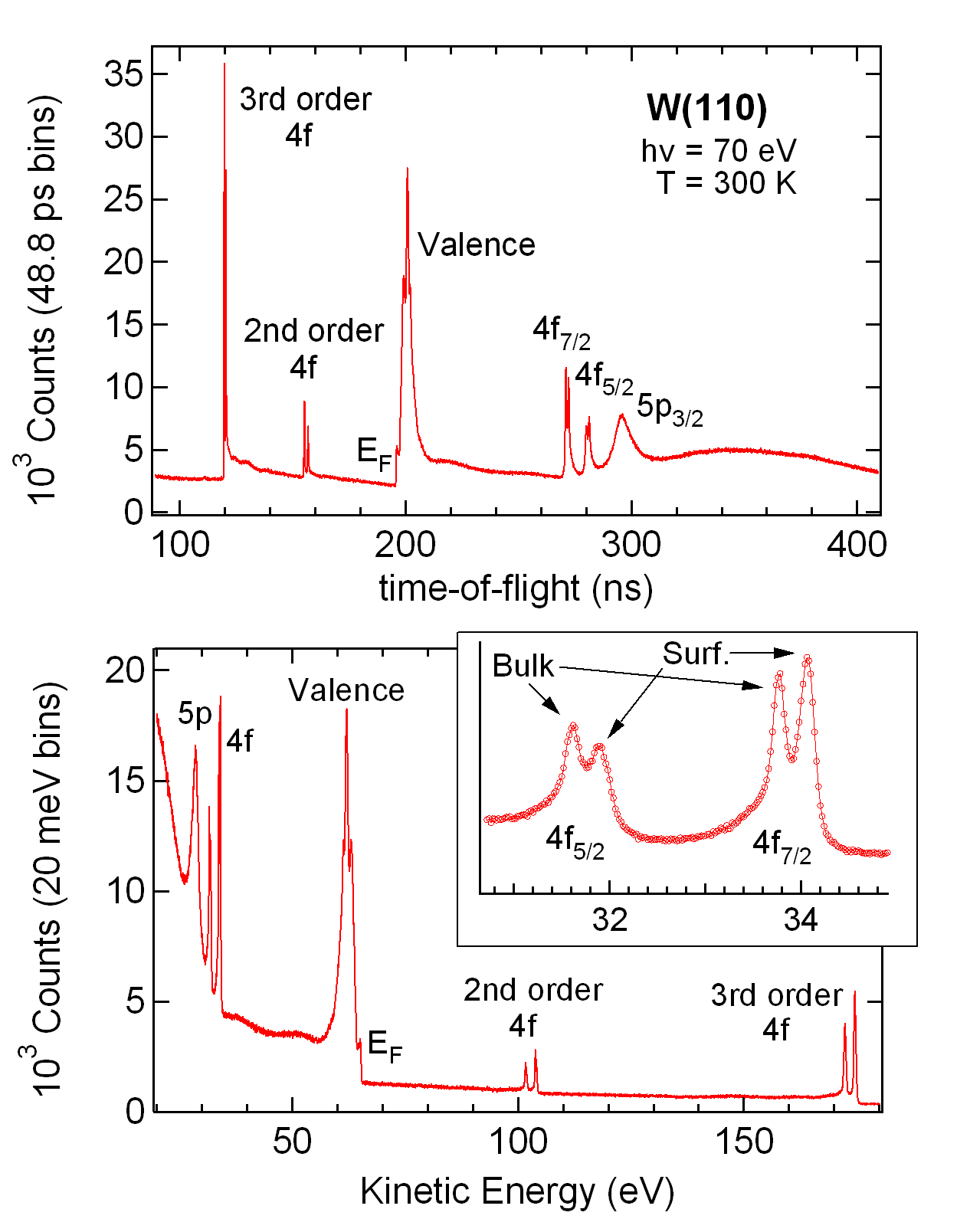}
\caption{\label{fig:sampspectra}(color online) Typical ``drift tube'' photoemission spectrum taken with straight path (M1), acquired in $<5$ minutes with a total count rate of 135 KHz.  The angular resolution in the free drift case, defined by the size of the detector and flight length, is $\pm0.4^{\circ}$.  The upper panel is the raw time spectrum, and the lower panel is converted to an energy axis.  The inset shows a close up of the W 4f core levels.  The splitting of the surface component illustrates a clean surface.}
\end{figure}

Fig.~\ref{fig:sampspectra} shows a typical W spectrum taken in M1 with all the electrodes grounded as a simple ballistic drift tube.
The upper panel shows the raw TOF spectrum, with $t=0$ aligned to the sharp photon `prompt' peak (not shown, section~\ref{sec:daq}), corrected for the $\sim$3 ns required for photons scattered from the sample to reach the detector.
Features of the spectrum are readily recognized, with the Fermi edge and intense valence band signal arriving at $\sim$ 200 ns, and core levels from 270-300 ns.
Intense peaks at shorter flight times are core levels excited by the presence of higher order light from the beamline.
The lower panel shows the same spectrum converted to an energy axis;\cite{Hemmers1998} the intensity scaling effect of the nonlinear conversion from equal time bins to equal energy bins (section~\ref{sec:TOFEEA}) is apparent as the signal from higher order light is now indeed less intense than that from first order light.
The features included in this spectrum allow convenient verification and calibration of the energy scale.
With the beamline set to $h\nu$=70 eV, each core level is separated by exactly 70 eV from its corresponding peak in the signal from the $h\nu$=140 and 210 eV light present.
Additionally, the binding energy of each core level, including the 4f splittings, are well known and serve as excellent benchmarks.
Overall, close agreement is found, and this procedure is also used with various sets of electrode voltages for flight energy retardation for calibration of the time-to-energy conversion.
In these cases, taking several spectra with incremental changes in photon energy also aids in the process, as described in Ref.~\onlinecite{Hemmers1998}.

The spectrum of Fig.~\ref{fig:sampspectra} illustrates the benefits of M1 operation and the TOF technique in general.
The entire spectrum is taken at once, which is often useful for rapid initial sample identification and characterization.
As an example, an expanded view of the 4f core levels is shown in the inset to the lower panel.
The strong surface splitting of each multiplet\cite{Riffe1989} is convenient for monitoring surface quality and cleanliness during an experiment, and rapid simultaneous monitoring of both the valence band and core levels can be useful during deposition of precision thickness overlayers.
Sensitivity is also good, showing clear features even from the extremely low flux of second order light.
Quick measurement of valence band peak dispersion as a function of emission angle is also useful for sample alignment. 

Note that in the example of Fig.~\ref{fig:sampspectra}, all the observed features are due to relatively high flight energy photoelectrons arriving within the first period after their originating photon pulse.
In other words, they correspond to being taken in the region corresponding to the first red line in Fig.~\ref{fig:overlap}, and are free of significant overlap from lower energy electrons.
However, recorded at a flight energy of 65 eV, the Fermi edge is fairly broad and is energy resolution limited by $\Delta E_t$ (equation~\ref{eq:dt}) to over 90 meV with $\Delta t$=137 ps.

\begin{figure} \includegraphics[width=8cm]{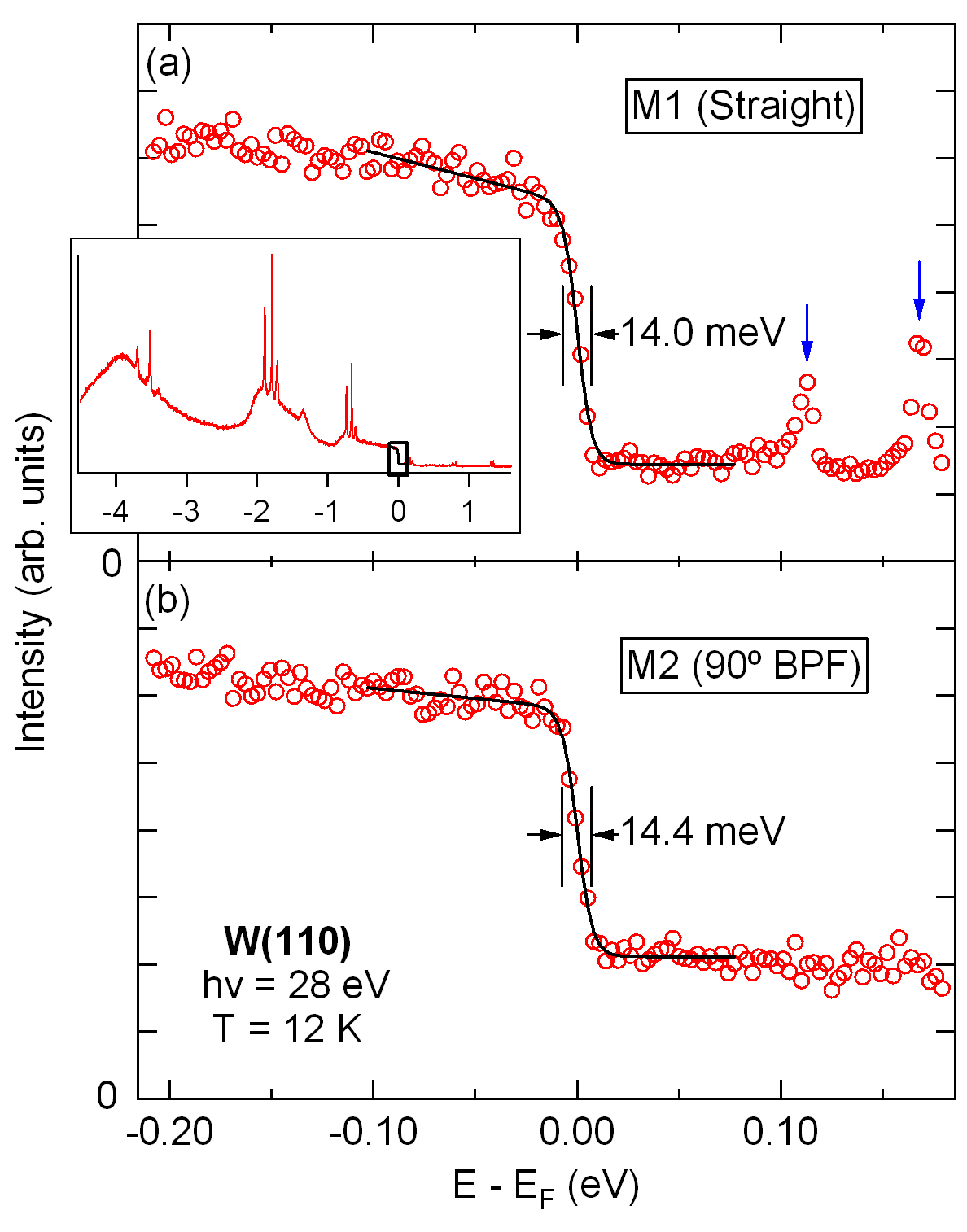}
\caption{\label{fig:enres}(color online) Photoemission spectra taken with the spin-integrating detector of the W Fermi edge showing total experimental energy resolution.  (a) Spectrum taken along straight path of M1 with $h\nu$=28 eV, $\sim$15 V retardation, and angular resolution $\pm0.6^{\circ}$.  Although the Fermi edge is sharp and recognizable, the spectrum badly overlaps with itself, resulting in the confusing full spectrum shown in the inset.  The box in the inset corresponds to the range expanded in the main panel.  The peaks under the vertical arrows mark signal from the 4f core levels coming from higher order light of pulses one period after those causing the signal at E$_F$.  (b) Same spectrum taken with BPF in M2 with $\sim$19 V retardation and angular resolution $\pm0.4^{\circ}$.  The energy resolution is similar to (a), but the spectrum is completely free of spectral overlap.}
\end{figure}

To resolve features near E$_F$ with higher resolution, we must use retardation for lower flight energies.
Fig.~\ref{fig:enres} (a) shows the W Fermi edge, measured in M1 with $h\nu$=28 eV and retarded to an average flight energy of $\sim$10.2 eV (energy axis referenced to E$_F$).
The measured edge is fit with a width of 14.0 meV; removing contributions from beamline resolution and temperature broadening ($\sim$ 10 meV and 4 meV, respectively) leaves a total $\Delta E$=9 meV for the spectrometer.
This is slightly larger than the $\Delta E_t$=5.9 meV expected from the 137 ps $\Delta t$,\cite{mstar} but is in quite reasonable agreement.
The small remaining contributions may be due to slight uncertainty in beamline resolution, small residual magnetic fields, minor sample surface imperfections, and slight detector misalignments leading to contributions to $\Delta t_{EEA}$.

Since the feature of interest is now at a low flight energy, it is overlapped with spectral intensity at significantly higher energies.
Indeed, this is the source of the sharp peaks just above E$_F$ marked by the vertical arrows; note that there are no such features in the actual photoemission spectrum 0.10 eV above E$_F$.
The inset to Fig.~\ref{fig:enres} (a) shows a wide view of the same spectrum.
The broad features from 0 to -4 eV are valence band features, while all the intense sharp features are due to overlapping, higher kinetic energy core level photoelectrons excited by higher order light which clearly distort the spectrum.
Lineshape analysis beyond the simple edge width extraction here is therefore not possible.

M2 operation with the 90$^{\circ}$ BPF removes these unwanted electrons for clean spectra.
Fig.~\ref{fig:enres} (b) shows the same spectrum, taken in M2 with an average flight energy of 6.6 eV.
Here the spin-integrating MCP1 was attached at the end of LS3 for a direct comparison of the two modes with the same detector.
A Fermi edge with a similar width of 14.4 meV was achieved even with the added complexity of the BPF and related timing aberrations.
Most importantly, the spectrum is completely free of overlap contamination: there is no sign of the peaks marked by arrows in panel (a).
Thus spectra taken at high energy resolution in M2 are representative of true spectra and more suitable for lineshape analysis.

It should be noted that the BPF in M2 operation introduces a strong transmission envelope function to recorded spectra; this is of course its purpose.
The electron optics of the TOF-EEA can be tuned to provide a fairly `square' transmission envelope, however measured spectra should be corrected for variations in the transmission function.\cite{Jozwiakthesis}
For all M2 spectra presented in the present paper, this has been performed by normalizing the recorded spectra by experimentally measured transmission functions.
These were obtained by recording spectra with identical TOF-EEA settings, but with higher photon energies to place known flat regions of the resulting photoemission spectrum within the bandpass.
In our current experience, this technique results in reliable spectra with quite good agreement with those taken in M1.
With further calibration of the TOF-EEA, the M2 transmission function can also be integrated out by sweeping the desired portion of the photoemission spectrum through the bandpass with appropriate control of lens voltages.

\subsection{\label{sec:spin}Spin resolution}

The well known Rashba spin-splitting of the Au(111) surface state\cite{LaShell1996,Hoesch2004} provides a convenient test subject for the polarimeter.\cite{Cacho2009}
The Au(111) surface was prepared by evaporating $\sim$15 monolayers of Au onto the clean W(110) substrate at room temperature.
M1 operation with the spin-integrating detector was useful for verifying substrate cleanliness as well as monitoring the Au growth.

\begin{figure} \includegraphics[width=8.5cm]{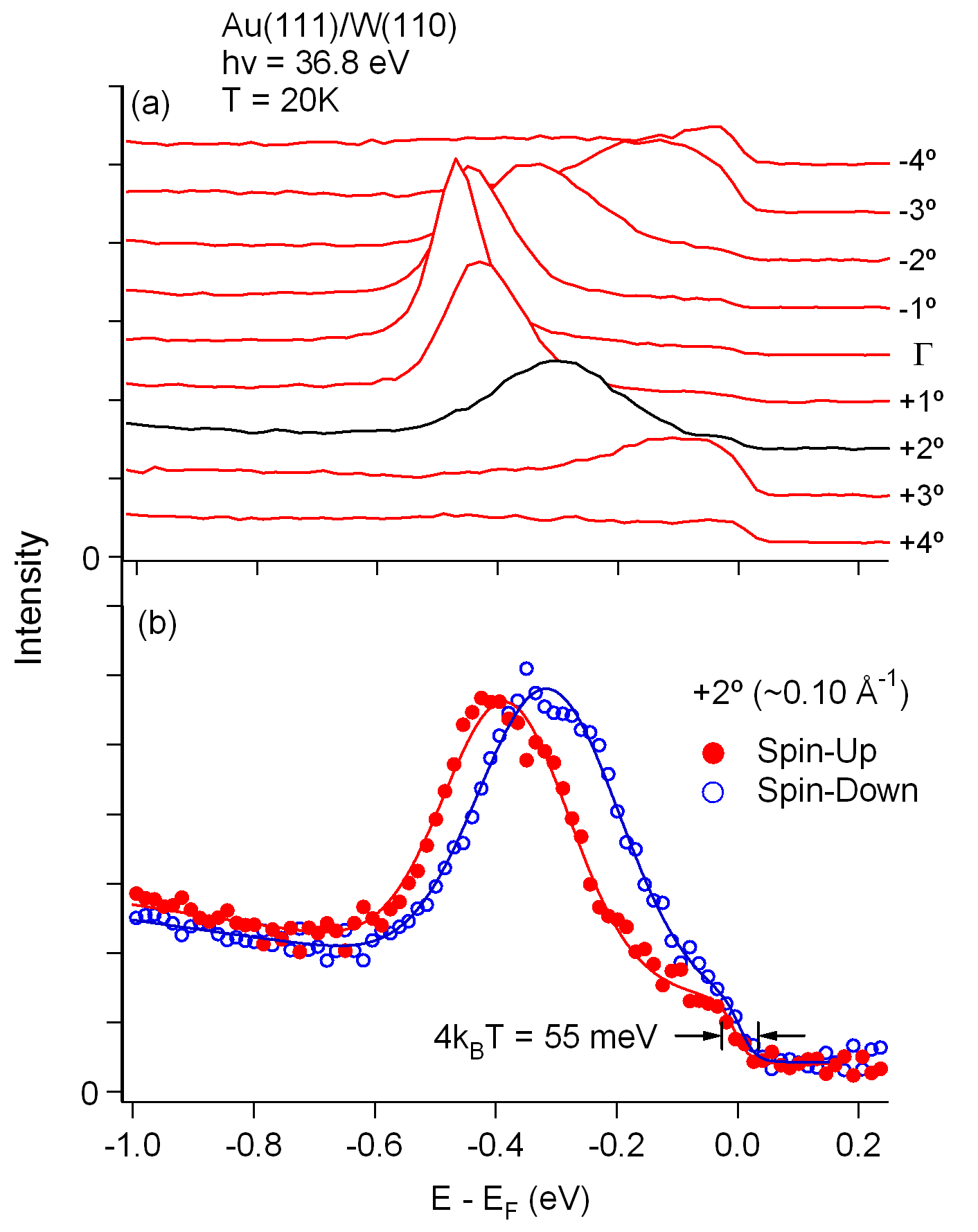}
\caption{\label{fig:AuSR}(color online) Spectra taken from Au(111) surface state.  (a) Spin-integrated M1 data as a function of emission angle along $\Gamma K$.  Spectra are vertically shifted for clarity.  (b) Spin-resolved M2 data, taken at the angle corresponding to black spectrum in (a), with angular resolution $\pm0.9^{\circ}$ and a total acquisition time of 3 hours.  Solid curves are fits to Gaussian peaks with a linear background and the Fermi function.}
\end{figure}

Fig.~\ref{fig:AuSR} (a) shows spin-integrated spectra as a function of emission angle along the Au(111) $\Gamma K$ direction taken in M1 with a sample temperature of 20 K.
The surface state is clearly seen as a parabolic dispersing peak; the splitting of the band is not resolved, likely because of the limited angular ($k$-space) resolution of $\Delta\theta$=0.8$^{\circ}$ ($\Delta k$=0.04\AA$^{-1}$).
The splitting should still be observable, however, with spin-resolution along the appropriate direction.
Panel (b) shows the spin-resolved spectrum at $+$2$^{\circ}$ taken with the polarimeter in M2.
Data were taken with the scattering target magnetization axis aligned perpendicular to the sample emission angle, i.e., with $\omega$ rotated 90$^{\circ}$ to align $y_t$ along $y_s$ (Fig.~\ref{fig:main}) and the Au sample rotated about the $y_s$ axis.
This makes the polarimeter sensitive to polarization along the sample's $k_y$ direction, perpendicular to the emission direction along $k_x$.
The resulting asymmetry in the collected spectra ($I_1$ and $I_2$) is converted into a measurement of $P_y(E)$ with equation~\ref{eqn:mainpol} using $S_{\textrm{eff}}$=0.12, as discussed in section~\ref{sec:SpinPol}.
The spin-up (spin-down) spectrum shown, $I_{\uparrow}$ ($I_{\downarrow}$), is extracted as $I_{\uparrow (\downarrow )}=1/2(I_1+I_2)(1\pm P_y)$.

The spin-resolved spectrum agrees quite well with the literature,\cite{Hoesch2004,Cacho2009} and the energy splitting of the spin-up and -down bands of $\sim$70 meV is in line with what is expected at the measured $k$ value ($k_x\approx 0.1$ \AA$^{-1}$).\cite{LaShell1996,Reinert2001}
The 4k$_B$T Fermi edge widths from fitting the spin-resolved spectra, as well as the spin-averaged spectrum (i.e. $1/2(I_1+I_2)$), are $\sim$55 meV, representing the total experimental energy resolution of the measurement.
This demonstrated $\Delta E \sim$55 meV in a spin-ARPES measurement of the Au(111) surface state is a significant improvement beyond the $\Delta E$=120 meV of the previous measurement by Hoesch \textit{et al.}\cite{Hoesch2004} taken with a Mott-HEA type spectrometer.\cite{Hoesch2002}
A more recent experiment by Cacho \textit{et al.}\cite{Cacho2009} performed with a Mott-TOF spectrometer with $h\nu$=6.2 eV laser excitation claims a $\Delta E$=28 meV, but this value was extracted from the much broader measured Fermi edge width at room temperature ($>$100 meV). 

Although the energy resolution currently demonstrated is an improvement beyond the most recent spin-ARPES measurements\cite{Hsieh2009,Hsieh2009a,Wells2009} and compares well (with improved angular resolution) with some of the highest energy resolution work published,\cite{Fedorov1998,Fedorov2002} it does not match the 14.4 meV capability of the spectrometer demonstrated with the spin-integrating detector (figure~\ref{fig:enres}).
The broadening in energy resolution with the polarimeter is possibly due to timing aberrations from incorrect tuning of the BPF focal point (see section~\ref{sec:TOFEEA}) for the spin-resolved measurements, from scattered electrons in the small entrance tube of the polarimeter, or from contributions of electrons inelastically scattered from the target.
These issues are under investigation with electron trajectory calculations, and an improved design of the polarimeter detector for reducing timing aberrations will be reported in a later publication.
Once these issues are dealt with, the polarimeter is expected to function with the same energy resolution as demonstrated by the spin-integrating detector above (Fig.~\ref{fig:enres}(b)), without further reduction in count rates or longer acquisition times.

\begin{figure} \includegraphics[width=7.5cm]{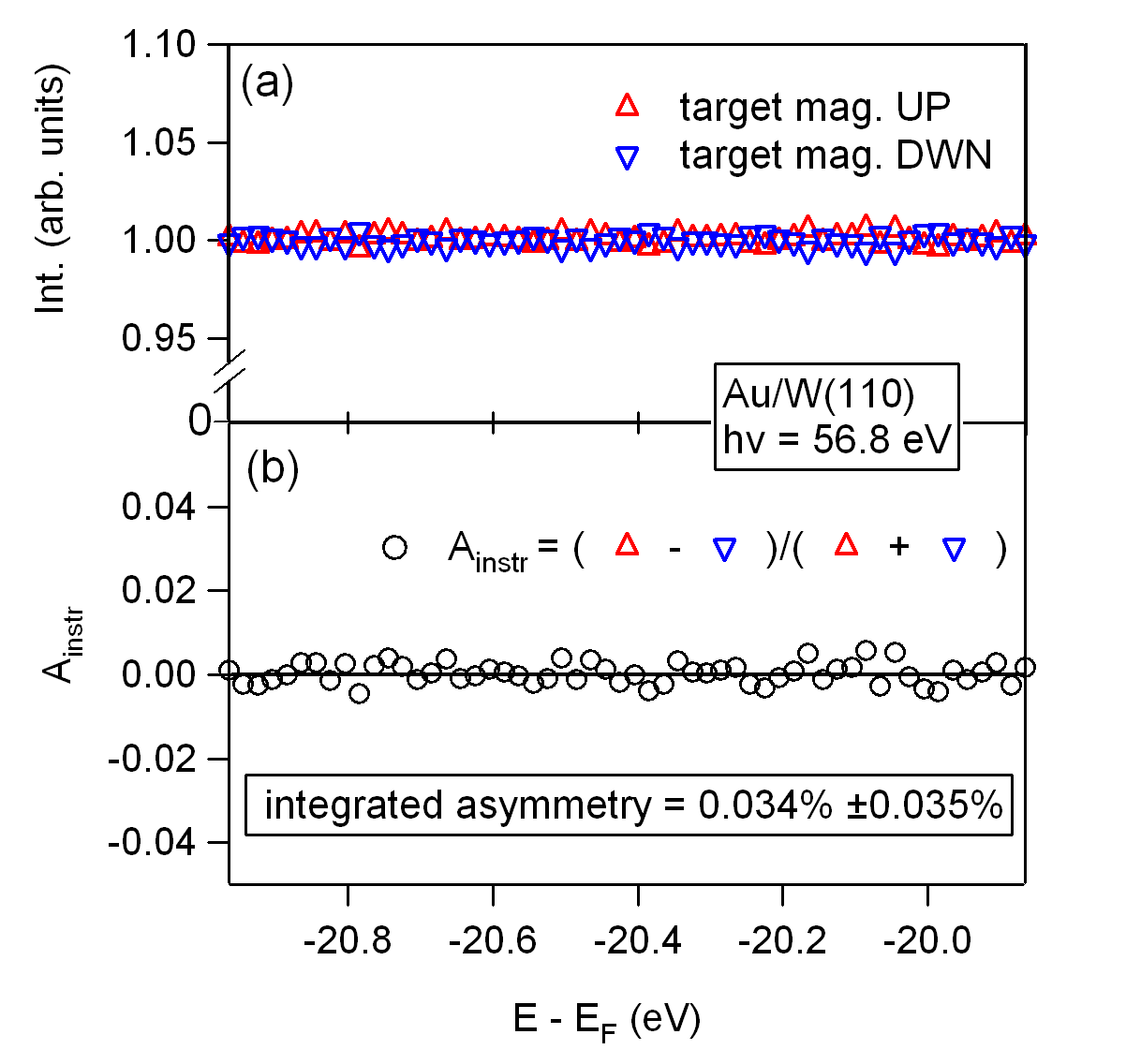}
\caption{\label{fig:nopol}(color online) The flat and featureless portion of the Au/W(110) photoemission spectrum corresponding to binding energy $\sim$20 eV.  (a) The intensity measured with the scattering target magnetized `up' (red upward triangles) and `down' (blue downward triangles).  The resulting asymmetry is shown in (b).  As no polarization is expected, this shows an extremely low $A_{instr}$.}
\end{figure}

As discussed in section~\ref{sec:SpinPol}, the $A_{\textrm{instr}}$ of a spectrometer is a significant concern for accurate polarization measurements.
A straightforward method for measuring the $A_{\textrm{instr}}$ is to analyze an unpolarized source of electrons.
The flat and featureless portion of the Au photoemission spectrum at binding energies between the valence bands and 4f core levels ($\sim$20 eV) is a convenient such benchmark.
The spectrometer voltages were left the same as for Fig~\ref{fig:AuSR} (b), but the photon energy was increased by 20 eV to measure this flat background signal under identical spectrometer conditions.
Figure~\ref{fig:nopol}(a) shows the resulting spectra with the polarimeter scattering target magnetized in both directions; the normalized asymmetry is shown in panel (b) and is defined as the spectrometer's $A_{\textrm{instr}}$ as no asymmetry due to spin polarization is expected.
Very little asymmetry is measured and deviations from zero seem only due to noise.
The total asymmetry integrated across the spectrum shown is 0.034\% $\pm$0.035\%.
This is far lower than Mott polarimeters (typical $\sim$10\%, as low as 2.6\%\cite{Cacho2009}) and other LEX polarimeter systems (3\%\cite{Bertacco1999}, 10-15\%\cite{Bertacco2002}), and is an order of magnitude lower than the lowest value reported in the literature to our knowledge (0.3\%\cite{Winkelmann2008}).
Thus error in polarization measurements due to instrumental asymmetry (equation~\ref{eq:instr}) is effectively removed, and double measurements with reversal of source polarization is not necessary.
LEX-polarimeters can be less sensitive to instrumental asymmetries than Mott polarimeters in general as they use a single detector for both intensity measurements.
The present design likely results in a particularly low instrumental asymmetry due to the high quality magnetic shielding, the low stray field from the scattering target, and the low stray field from the magnetizing coil design.
Further, the instrumental asymmetries of LEX-polarimeters are in general less sensitive to beam alignment than Mott polarimeters as the $S_{\textrm{eff}}$ does not depend significantly on the incident angle or target beam spot position within a few degrees or mm, respectively.\cite{Bertacco1999}
In the present measurements, the photon beam spot is simply aligned with respect to the spin-TOF analyzer by maximizing the total count rate, which is insensitive to position through more than 100\% of the beam size.
As no further precise alignment is used, through significant experience it is clear that the presented  instrumental asymmetry is not at all dependent on beam alignment through this range of positions and likely much more.

\section{\label{sec:sum}Summary}

We have presented a new spectrometer for spin-resolved ARPES, integrating a newly designed LEX-polarimeter into a flexible TOF energy analyzer.
LEX-polarimeters offer increases in detection efficiency of up to 100 times that of Mott polarimeters, while TOF based spin-resolved systems offer at least an order of magnitude efficiency increase over the wider-spread hemispherical analyzers.
This particular combination has not, to our knowledge, been previously developed.

The spectrometer includes two modes of operation with distinct flight paths for both rapid, full spectrum, spin-integrated measurements, and more narrow, high-resolution spin-resolved measurements.
The inclusion of a 90$^{\circ}$ dispersive element in the latter path allows for lower flight energies and higher energy resolutions than would otherwise be possible with most synchrotron light sources.
Energy resolutions at $E_F$ as low as 10 meV with photon energies of 28 eV have been demonstrated.
The polarimeter allows flexible orientation of the spin-analysis axis anywhere within the transverse plane and allows straightforward use of a wide variety of scattering target systems.
An instrumental asymmetry of $<0.04$\% is directly measured.
Full operation of the spectrometer is demonstrated through the acquisition of spin-ARPES of the Rashba spin-split Au(111) surface state. 
Complete realization of the potential of this instrument can provide much improved efficiency and resolution for the spin-ARPES experiments in current demand.  

\textit{Note added.}
During the submission process of this manuscript, work to further improve the energy resolution achieved with the spin polarimeter proceeded as discussed in Sec.~\ref{sec:spin}.  A more complete graphite coating of the electron-optical surfaces within the polarimeter was performed, and extra apertures were installed for removing possible electrons that scatter from the optics upstream of the polarimeter and those that scatter inelastically from the target far from the specular geometry.  A total experimental energy resolution of 20 meV was recently demonstrated with the spin-polarimeter in M2 mode of the spin-TOF spectrometer at beamline 12.0.1 of the ALS during 2-bunch mode with very similar count rates as in the present paper.  Removing contributions from beamline resolution and temperature broadening in this experiment ($\sim$ 15 meV and 6 meV, respectively) leaves a total of $\Delta E$$\sim$12 meV for the spectrometer.  These results will be further discussed in a later publication.

\begin{acknowledgments}
We would like to thank Hans Siegmann, Neville Smith, and Z.X. Shen for providing the inspiration that began this project.
We thank B. Sinkovic and Y.L. Chen for continually providing their experimental experience and expertise. 
We also thank S. Chourou for early help with electron-optical calculations, simulations, and explanations, J.S. Pepper, S. DiMaggio, and A. Williams for invaluable machining and technical support, and C.G. Hwang, D.R. Garcia, D.A. Siegel, and S.D. Lounis for experiment assistance.
One of us (A. Lanzara) would like to thank the University of California, Berkeley for partly supporting this project through faculty start-up funds.
The work was additionally supported by the Director, Office of Science, Office of Basic Energy Sciences, Materials Sciences and Engineering Division, of the U.S. Department of Energy under Contract No. DE-AC02-05CH11231.
The work was also supported by, and performed at the Advanced Light Source, Lawrence Berkeley National Laboratory, which is supported by the Director, Office of Science, Office of Basic Energy Sciences, of the U.S. Department of Energy under Contract No. DE-AC02-05CH11231.
\end{acknowledgments}


\begin{thebibliography}{10}
\bibitem{woverde}
This can be seen as follows. The HEA only accepts electrons within an energy
  range of $\Delta E$, and must sequentially scan this range through $W$. The
  TOF simultaneously accepts electrons through the entire $W$, and thus is
  $W/\Delta E$ times more efficient. This simple comparison assumes a constant
  $\Delta E$ through $W$ for the TOF; as $\Delta E$ is dependent on the flight
  energy, $E$, this is approximately valid for small ranges of $W$ compared to
  $E$.
\bibitem{Hamamatsu4294}
MCPs were removed from a Hamamatsu F4294-09 detector assembly and inserted into
  a custom assembly.
\end{thebibliography}
\end{document}